\documentclass[11pt,aps,prd,showpacs,nofootinbib,superscriptaddress,preprint,preprintnumbers]{revtex4}

\pdfoutput=1

\usepackage[usenames,dvipsnames]{color}
\usepackage{graphicx}
\usepackage{amsmath,amsfonts,amssymb,dsfont,bm,bbm,dcolumn,float,ifthen,mathrsfs}
\usepackage{pstricks,feyn,verbatim}
\usepackage{url,hyperref}

\usepackage{slashed,xcolor,multirow,array,rotating}
\usepackage{fancybox,epstopdf}

\newcommand{\GeV}      {~\mathrm{GeV}}
\newcommand{\TeV}      {~\mathrm{TeV}}

\newcommand{\pb}      {~\mathrm{pb}}
\newcommand{\fb}      {~\mathrm{fb}}

\newcommand{\ba}{\begin{array}}
\newcommand{\ea}{\end{array}}
\newcommand{\beqn}{\begin{eqnarray}}
\newcommand{\eeqn}{\end{eqnarray}}
\newcommand{\beqs}{\begin{subequations}}
\newcommand{\eeqs}{\end{subequations}}
\newcommand{\be}{\begin{equation}}
\newcommand{\ee}{\end{equation}}
\newcommand{\non}{\nonumber \\}

\newcommand{\mathsym}[1]{{}}

\def\mL{\mathcal{L}}

\def\mO{\mathcal{O}}

\def\hf{\frac{1}{2}}

\usepackage{ulem,fancyvrb}
\usepackage{xcolor} 
\begin{document}

\title{LHC searches for the CP-odd Higgs by the jet substructure analysis}

\begin{flushright}
ADP-14-28-T888\\
\end{flushright}

\author{Ning Chen}
\email{chenning@ustc.edu.cn}
\affiliation{Department of Modern Physics, University of Science and Technology of China, Hefei, Anhui, 230026, China}
\author{Jinmian Li}
\email{jinmian.li@adelaide.edu.au}
\affiliation{ARC Centre of Excellence for Particle Physics at the Terascale, School of Chemistry and Physics, University of Adelaide, Adelaide, SA 5005, Australia}
\author{Yandong Liu}
\email{ydliu@pku.edu.cn}
\affiliation{Department of Physics and State Key Laboratory of Nuclear Physics and Technology, Peking University, Beijing 100871, China}
\author{Zuowei Liu}
\email{zuoweiliu@tsinghua.edu.cn}
\affiliation{Center for High Energy Physics, Tsinghua University, Beijing, 100084, China}

\begin{abstract}
The LHC searches for the CP-odd Higgs boson $A$ is studied (with masses from $300\, \GeV$ to $1\,\TeV$) in the context of the general two-Higgs-doublet model. With the discovery of the $125\,\GeV$ Higgs boson at the LHC, we highlight one promising discovery channel of $A\to hZ$. This channel can become significant for heavy CP-odd Higgs boson after the global signal fitting to the $125\,\GeV$ Higgs boson in the general two-Higgs-doublet model. It is particularly interesting in the scenario where two CP-even Higgs bosons in the two-Higgs-doublet model have the common mass of $125\,\GeV$. Since the final states involve a Standard-Model-like Higgs boson, we apply the jet substructure analysis of tagging the fat Higgs jet in order to eliminate the Standard Model background sufficiently. After performing the kinematic cuts, we present the LHC search sensitivities for the CP-odd Higgs boson with mass up to $1\,\TeV$ via this channel. 

\end{abstract}


\maketitle

%###################################################################

\section{Introduction}
\label{section:intro}

The study of the Higgs mechanism~\cite{Higgs:1964ia, Higgs:1964pj, Englert:1964et} has become more interesting and important since the discovery of the $125\,\GeV$ Higgs boson at the LHC $7\oplus 8\,\TeV$ runs. The properties of the $125\,\GeV$ Higgs boson, such as the coupling strengths with Standard Model (SM) fermions and gauge bosons~\cite{Peskin:2012we}, its spin and parity~\cite{Djouadi:2013yb}, and the exotic decay channels~\cite{Curtin:2013fra}, will be further measured in the next LHC runs and the future high energy colliders. From various motivations, the SM Higgs mechanism is far from being complete. New physics models beyond the SM (BSM) are proposed to address different questions, which typically contain new states in the spectrum. In many of them, the electroweak symmetry breaking (EWSB) is due to the extended Higgs sector. Examples include the minimal supersymmetric extension of the SM (MSSM)~\cite{Dimopoulos:1981zb}, the twin Higgs models~\cite{Chacko:2005vw}, and the composite Higgs models~\cite{Mrazek:2011iu}. The future experimental searches for the new degrees of freedom in the spectra provide direct avenues for revealing the underneath new physics.

A very widely studied scenario beyond the minimal one-doublet setup is the two-Higgs-doublet model (2HDM), which is the low-energy descriptions of the scalar sectors in many new physics models. A recent review of the phenomenology in the context of the general 2HDM can be found in Ref.~\cite{Branco:2011iw}. Refs.~\cite{Craig:2012vn, Craig:2012pu, Coleppa:2013dya, Craig:2013hca, Coleppa:2013xfa,Carena:2013ooa, Chen:2013emb, Baglio:2014nea, Coleppa:2014hxa, Dumont:2014wha, Dorsch:2014qja, Hespel:2014sla, Barger:2014qva, Fontes:2014xva, Coleppa:2014cca, Grzadkowski:2014ada} studied the 2HDM phenomenology at the LHC in light of the Higgs discovery. The scalar spectrum in the 2HDM contains five states, namely, two neutral CP-even Higgs bosons $(h\,,H)$, one neutral CP-odd Higgs boson $A$, and two charged Higgs bosons $H^\pm$. Often, one would interpret the lighter CP-even Higgs boson $h$ as the one discovered at the LHC. In the context of the general 2HDM, each Higgs boson mass is actually free parameter before applying any constraint. Therefore, it is likely that two or more states become degenerate in mass~\cite{Gunion:2012he}.

Within the framework of the 2HDM, we study the future LHC searches for the CP-odd Higgs boson $A$ at the $14\,\TeV$ run. The previous experimental searches often focus on the benchmark models in the MSSM, which has a type-II 2HDM Yukawa couplings. Thus, the interesting final states to be looked for are the $A\to \bar b b$~\cite{Aaltonen:2012zh,Chatrchyan:2013qga} and $A\to \tau^+ \tau^-$~\cite{Schael:2006cr,Abazov:2008hu,Aaltonen:2009vf, Abazov:2011up,Chatrchyan:2012vp, Aad:2012cfr, Aad:2014vgg} since the relevant Yukawa couplings are likely to be significantly enhanced. Different from the existing experimental search modes, we focus on the decay channel of $A \to hZ$. The previous studies to this search channel at the LHC include Refs.~\cite{Coleppa:2013dya,Coleppa:2013xfa, Coleppa:2014hxa}, where the final states of $\bar b b \ell^+ \ell^- $, $\tau^+ \tau^- \ell^+ \ell^-$, and $ZZZ$ were studied. Also, an experimental analysis of this search channel with multiple lepton and photon final states was carried out at the LHC $8\,\TeV$ run~\cite{CMS:2013eua}. Here, in our analysis, we will focus on the $\bar b b\ell^+ \ell^-$ final state coming from the decay channel of $A\to hZ$. In this case, the final states involve a SM-like Higgs boson with mass of $125\,\GeV$. Therefore, the jet substructure method of tagging the boosted Higgs jet can be potentially instrumental for this particular channel in our study. The method of tagging the boosted Higgs jets was suggested in Ref.~\cite{Butterworth:2008iy, Butterworth:2008tr}, in which the discovery potential of the SM Higgs boson via the $hV$-associated production channel at the LHC was investigated. Later, this procedure was widely adopted in searches for new resonances with a SM-like Higgs boson as their decay final states~\cite{Butterworth:2008tr, Carena:2010ev, Yang:2011jk, Kang:2013rj} and in studies of the SM-like Higgs boson properties at the LHC~\cite{Plehn:2009rk, Godbole:2013saa, Godbole:2014cfa}.

This paper is organized as follows. In Sec.~\ref{section:2HDM}, we give a brief review on the CP-odd Higgs boson $A$ in the context of the general 2HDM. We list its coupling terms, with the emphasis on the derivative couplings of $AhZ$ and $AHZ$. In Sec.~\ref{section:AProdDecay}, we evaluate the productions and decays of the CP-odd Higgs boson $A$ in the context of the general 2HDM. We show that the decay mode of $A\to hZ$ can be sizable for the future LHC searches at the $14\,\TeV$ runs, given the current global fit to the 2HDM parameters. We also show for the degenerate Higgs scenario of $M_h = M_H = 125 \,\GeV$, the decay modes of $A\to hZ/HZ$ are typically dominant over all other decay modes into the SM final states. Hence, such a mode can be regarded as the leading one for the future searches for the CP-odd Higgs boson in this special case. In Sec.~\ref{section:AhZ}, the analysis of LHC searches for the CP-odd Higgs boson via the $A\to hZ$ final states is provided. In order to eliminate the SM background sufficiently, we apply the jet substructure method developed in Ref.~\cite{Butterworth:2008iy} to tag the fat Higgs jet directly with the Cambridge/Aachen (C/A) jet algorithm~\cite{Dokshitzer:1997in,Wobisch:1998wt}. Optimizations to the jet substructure methods and kinematic cuts for the signal processes are presented. The LHC search potential to the $A\to hZ$ decay channel at different phases of the upcoming runs at $14\,\TeV$ are also shown. The conclusions are given in Sec.~\ref{section:conclusion}.

%###################################################################

\section{The CP-odd Higgs Boson in The 2HDM}
\label{section:2HDM}

\subsection{The CP-odd Higgs boson mass}

The most general 2HDM Higgs potential that is CP-conserving contains two mass terms plus seven more quartic coupling terms. For simplicity, we consider the soft breaking of a discrete $\mathbb{Z}_{2}$ symmetry, under which the two Higgs doublets transform as $(\Phi_1\,,\Phi_2)\to (\Phi_1\,, - \Phi_2)$. The simplified 2HDM potential is expressed as
\beqn\label{eq:2HDM_potential}
V(\Phi_1\,,\Phi_2)&=&m_{11}^2|\Phi_1|^2+m_{22}^2|\Phi_2|^2-m_{12}^2(\Phi_1^\dag\Phi_2+H.c.)+\hf\lambda_1 |\Phi_{1}|^{4} +\hf\lambda_2|\Phi_{2}|^{4}\non
&+&\lambda_3|\Phi_1|^2 |\Phi_2|^2+\lambda_4 |\Phi_1^\dag \Phi_2|^2+\hf\lambda_5 \Big[ (\Phi_1^\dag\Phi_2) (\Phi_1^\dag\Phi_2)+H.c.\Big]\,,
\eeqn
where all parameters are real. The two Higgs doublets $\Phi_1$ and $\Phi_2$ pick up vacuum expectation values (VEVs) to trigger the EWSB
\beqn\label{eq:2HDM_vevs}
&&\langle \Phi_{1} \rangle =\frac{1}{\sqrt{2}}\left( \begin{array}{c} 0 \\ v_{1} \\ \end{array}  \right)\qquad  \langle \Phi_{2} \rangle =\frac{1}{\sqrt{2}}\left( \begin{array}{c} 0 \\ v_{2} \\ \end{array}  \right)\,,
\eeqn
and one often parametrizes the ratio of the two Higgs VEVs as
\beqn\label{eq:tb}
t_\beta&\equiv&\tan\beta\equiv \frac{v_2}{v_1}\,.
\eeqn
Expressing the two Higgs doublets in component, we have
\beqn\label{eq:2HDM_doublets}
\Phi_i&=&\left( \ba{c} \pi_i^+ \\ ( v_i + h_i + i \pi_i^0 )/\sqrt{2}   \ea \right)\,,~~~i=1\,,2\,.
\eeqn
Three of the eight components are Nambu-Goldstone bosons giving rise to the electroweak gauge boson masses, with the remaining five components as the physical Higgs bosons: two CP-even Higgs bosons, $h$ and $H$, one CP-odd Higgs boson $A$, and the charged Higgs bosons $H^\pm$. The CP-odd Higgs boson $A$ is a linear combination of the two imaginary components $\pi_i^0$ in the doublets: $A= - s_\beta \pi_1^0 + c_\beta \pi_2^0$, whereas the orthogonal linear combination of $G= c_\beta \pi_1^0 + s_\beta \pi_2^0$ corresponds to the Nambu-Goldstone mode to be eaten by the $Z$ boson. By extracting the relevant terms in the 2HDM potential (\ref{eq:2HDM_potential}), the CP-odd Higgs boson mass square is given by
\beqn\label{eq:MA}
M_A^2&=& ( m_{12}^2 - \lambda_5 v_1 v_2  ) ( t_\beta + 1/t_\beta)\,. 
\eeqn

%&&&&&&&&&&&&&&&&&&&&&&&&&&&&&&&&&&&&&&&&&&&&&&&

\subsection{The couplings of the CP-odd Higgs boson}

\begin{table}[htb]
\begin{center}
\begin{tabular}{c|c|c}
\hline
%\hline
 & 2HDM-I & 2HDM-II    \\\hline\hline  
 $\xi_{A}^{u}$  & $1/t_{\beta}$ & $1/t_{\beta}$  \\
 $\xi_{A}^{d}$  & $-1/t_{\beta}$ & $t_{\beta}$  \\
 $\xi_{A}^{\ell}$  & $-1/t_{\beta}$ & $t_{\beta}$ \\
 \hline
\end{tabular}
\caption{The Yukawa couplings of the SM quarks and charged leptons to the CP-odd Higgs boson $A$ in the 2HDM-I and 2HDM-II. }\label{tab:A_yuk}
\end{center}
\end{table}

At the tree level, the CP-odd Higgs boson $A$ couples to the SM fermions through the Yukawa coupling terms
\beqn\label{eq:AYukawa}
-\mL_Y^A&=&-i\sum_f  \frac{m_f}{v} \xi_A^f \bar f \gamma_5 f A\,,
\eeqn
where $f$ is the SM fermion, $m_f$ is the SM fermion mass, and $v=\sqrt{v_1^2 + v_2^2}= (\sqrt{2} G_F)^{-1/2}=246\,\GeV$. The relevant coupling strengths of $\xi_A^f$ are listed in Table.~\ref{tab:A_yuk} for the 2HDM-I and 2HDM-II cases. Details of the Yukawa setups in the 2HDM-I and 2HDM-II can be found in Ref.~\cite{Branco:2011iw}. The loop induced couplings such as $Agg$ and $A\gamma\gamma$ are also correlated with the Yukawa coupling strengths of $\xi_A^f$. Since we limit our discussions in the SM content expect for the scalar part, we do not consider the CP-odd Higgs couplings with supersymmetric particles such as charginos, neutralinos, and sfermions. In addition, there are relevant derivative couplings of the CP-odd Higgs boson $A$ with the $Z$ boson and the CP-even Higgs bosons $(h\,,H)$, which arise from the kinematic terms of $|D\Phi_{i}|^{2}$. The couplings of $AhZ$ and $AHZ$ read
\beqn\label{eq:AhZ_coup}
&\sim&\hf(g W^{3}-g_{Y}B)\cdot\Big[ h_i(\partial\pi_i^0 )-\pi_i^0 (\partial h_i )\Big]\non
&\Rightarrow&\frac{g}{2c_{W}}Z\cdot \Big\{ c_{\alpha-\beta} [ h(\partial A)-A(\partial h)  ]+ s_{\alpha-\beta} [ H(\partial A)-A(\partial H) ] \Big\}\,,
\eeqn
where we express them in terms of the mass eigenstates, $c_{\alpha-\beta}\equiv \cos(\alpha-\beta)$, and $s_{\alpha-\beta} \equiv \sin(\alpha - \beta)$. Here $\alpha$ represents the mixing angle of the CP-even Higgs bosons. In many cases, one would regard the lighter CP-even Higgs boson $h$ as the $125\,\GeV$ Higgs boson discovered at the LHC, while others are regarded as heavier scalars to be searched for in the upcoming LHC runs. This is generally true for the Higgs spectrum in the MSSM. In the context of the general 2HDM, we also consider the degenerate Higgs scenario with $M_h = M_H = 125\,\GeV$. The CP-odd Higgs boson $A$ can also decay to the final states of $H^\pm W^\mp$ due to the derivative coupling terms of $A H^\pm W^\mp$ in the 2HDM kinematic terms. In our study here, we will always take the heavy mass input for $M_{H^\pm}$ and the decay modes of $A\to H^\pm W^\mp$ will not be addressed. The searches for this decay mode was recently studied in Ref.~\cite{Coleppa:2014cca}.

At the end of this section, we mention the constraints on the 2HDM parameters in light of the $125\,\GeV$ Higgs boson discovery at the LHC. In studies of the 2HDM, it is often assumed that the lightest CP-even Higgs boson $h$ in the spectrum corresponds to the $125\,\GeV$ Higgs boson discovered at the LHC $7\oplus 8\,\TeV$ runs. Under this assumption, one can perform a global fit to the signal strengths of $h$ based on a particular 2HDM setup. Only two parameters $(\alpha\,,\beta)$ are relevant for determining the gauge couplings of $g_{hVV}$ and the Yukawa couplings of $g_{hff}$. Details of such global fits can be found in Refs.~\cite{Craig:2013hca, Barger:2013ofa}. Given that the current signals in various decay channels are generally close to the SM Higgs boson predictions, the global fits to the allowed 2HDM parameter regions on $(\alpha\,,\beta)$ are consistent to the so-called ``alignment limit'' where $c_{\beta - \alpha} \to 0$. Consequently, one has $g_{hVV} \to g_{hVV}^{({\rm SM})}$ and $g_{hff} \to g_{hff}^{({\rm SM})}$ in this limit. Due to different Yukawa coupling patterns, the allowed regions of $c_{\beta-\alpha}$ are typically $\sim\mO(0.1)$ for the 2HDM-I case, and are more stringently constrained to be $\sim \mO(0.01)$ in the 2HDM-II case. In the analysis below, we take the following alignment parameter sets
\beqn\label{eq:align_para}
&&{\rm 2HDM-I}:c_{\beta-\alpha}=0.2\,,~~{\rm 2HDM-II}:c_{\beta-\alpha}=-0.02\,,
\eeqn
when we take $h$ to be the only state with mass of $125\,\GeV$. Since the relevant coupling terms given in Eq. (\ref{eq:AhZ_coup}) depends on the angle $\alpha-\beta$, this suggests the partial width of $\Gamma[A\to hZ]$ is suppressed due to the smallness of $c_{\beta-\alpha}$. However, for the larger $M_A$ region, this decay mode is likely to dominate over the fermionic decay modes, such as $A\to \bar t  t$. Besides, we shall also consider the degenerate Higgs scenario with $M_h=M_H = 125 \,\GeV $ in the 2HDM spectrum, where one cannot distinguish the decay modes of $A\to hZ$ and $A\to HZ$. Under this case, the combined decay widths of $\Gamma[A\to h/H + Z]$ should be considered for the LHC analysis, which is no longer suppressed by the small $c_{\alpha-\beta}$ parameter, and thus the partial decay widths of $\Gamma[A\to h/H+Z]$ are generally dominant over all others for the CP-odd Higgs boson. In what follows, we will always use $A\to hZ$ for both the $M_h = 125\,\GeV$ scenario and the $M_h = M_H = 125\,\GeV$ scenario.

%###################################################################

\section{The Productions and Decays of The CP-odd Higgs Boson $A$}
\label{section:AProdDecay}

\subsection{The productions of $A$}

\begin{figure}
\centering
\includegraphics[width=5cm,height=2.5cm]{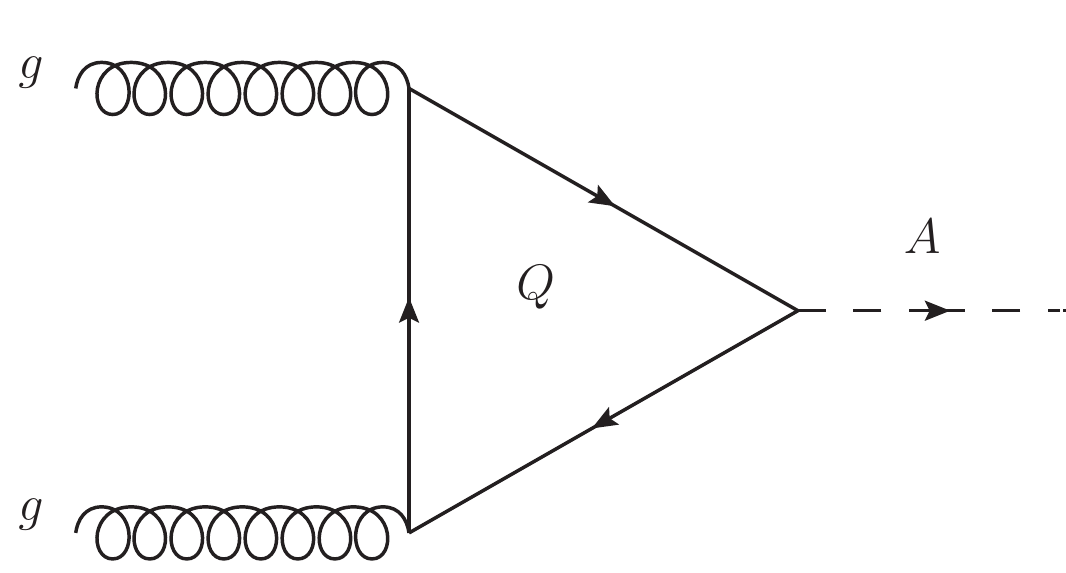}\\
\includegraphics[width=4.2cm,height=2.5cm]{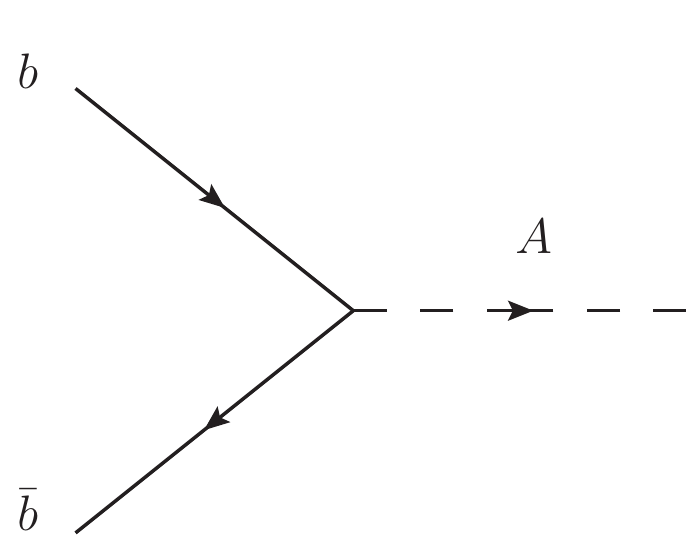}
\includegraphics[width=4.2cm,height=2.5cm]{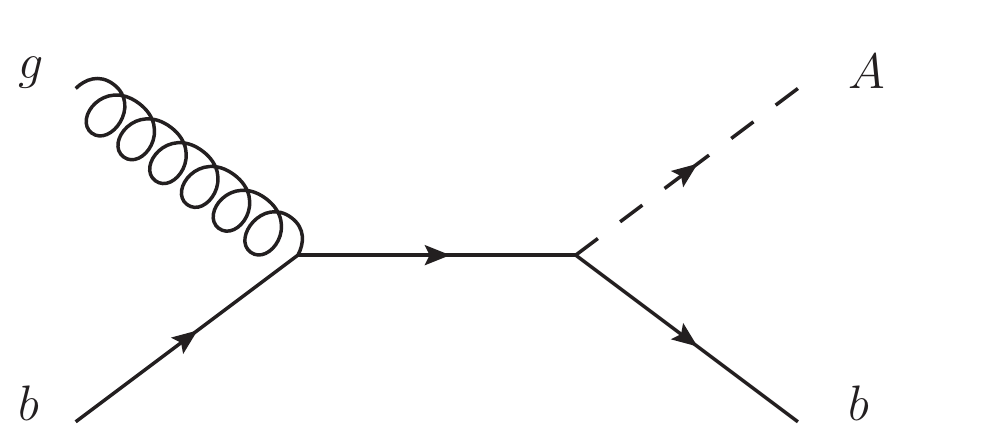}
\includegraphics[width=4.2cm,height=2.5cm]{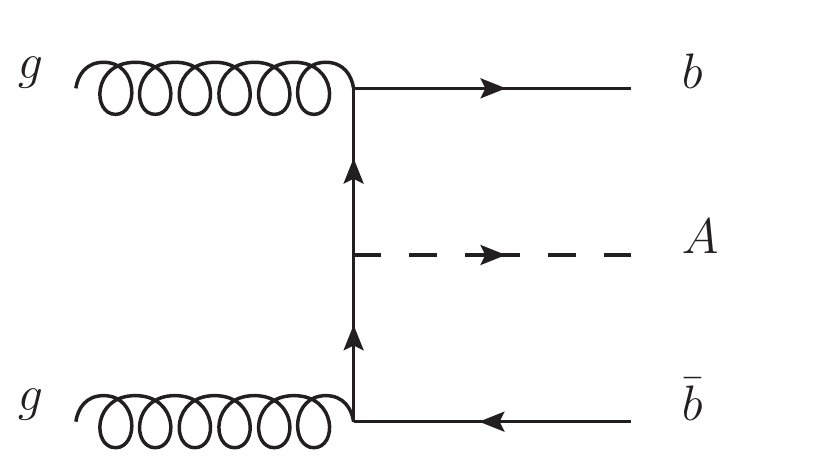}
\caption{\label{fig:Aprod} The Feynman diagrams for the production channels of the CP-odd Higgs boson $A$.}
\end{figure}

\begin{figure}
\centering
\includegraphics[width=7cm,height=4cm]{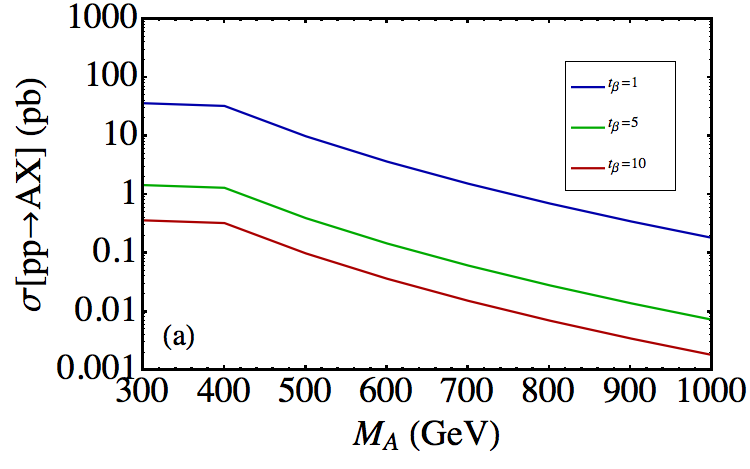}
\includegraphics[width=7cm,height=4cm]{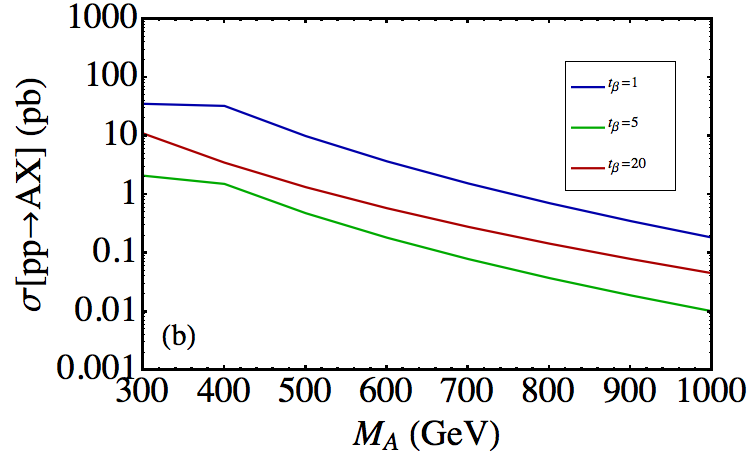}
\caption{\label{fig:pptoA} The inclusive production cross section $\sigma[pp\to AX]$ for $M_A \in (300\,\GeV \,, 1\,\TeV)$ at the LHC $14\,\TeV$ runs. Left: 2HDM-I, with inputs of $t_\beta=1$ (blue), $t_\beta=5$ (green), and $t_\beta=10$ (red). Right: 2HDM-II, with inputs of $t_\beta=1$ (blue), $t_\beta=5$ (green), and $t_\beta=20$ (red).}
\end{figure}

The CP-odd Higgs boson $A$ can be produced from both the gluon fusion and the bottom quark annihilation processes~\cite{Djouadi:2005gi, Djouadi:2005gj}. The relevant Feynman diagrams for these processes are depicted in Fig.~\ref{fig:Aprod}. At leading order, the partonic production cross section of $\hat \sigma(gg\to A)$ is related to the gluonic partial decay width as follows
\beqs
\beqn
\hat \sigma(gg\to A)&=&\frac{\pi^2}{8 M_A} \Gamma[A\to gg] \delta(\hat s - M_A^2)\,,\label{eq:ggA}\\
 \Gamma[A\to gg]&=&\frac{G_F \alpha_s^2 M_A^3}{64\sqrt{2} \pi^3 } \Big| \sum_q \xi_A^q A_{1/2}^A(\tau_q)   \Big|^2\,,\label{eq:Atogg}
\eeqn
\eeqs
with $\tau_{q}\equiv M_A^{2}/(4 m_{q}^{2})$ and $\xi_A^q$ being the Yukawa couplings given in Table.~\ref{tab:A_yuk}. Here $A_{1/2}^A(\tau)$ is the fermionic loop factor for the pseudoscalar. In the heavy quark mass limit of $m_q\gg M_A$, this loop factor reaches the asymptotic value of $ A_{1/2}^A(\tau)\to 2$, while it approaches zero in the chiral limit of $m_q\ll M_A$. For the 2HDM-I case, the dominant contribution to the gluon fusion process is always the top-quark loop; for the 2HDM-II case, however, the contribution from the bottom quark loop can become comparable to the top quark loop with the large $t_\beta$ inputs due to the different $t_\beta$ dependence in Yukawa couplings, as shown in Table.~\ref{tab:A_yuk}. Since we have $\xi_A^u=1/t_\beta$ in both 2HDM-I and 2HDM-II cases, the top quark loop in the gluon fusion process will be suppressed for the larger $t_\beta$ inputs. The bottom quark associated processes are controlled by the Yukawa coupling of $\xi_A^d$, which reads $-1/t_\beta$ in 2HDM-I and $t_\beta$ in 2HDM-II. Therefore, the contributions from these processes become sizable in the 2HDM-II with the large $t_\beta$ input. In practice, we evaluate the production cross sections for these processes by {\tt SusHi}~\cite{Harlander:2012pb}. The inclusive production cross sections of $pp\to AX$ are shown in Fig.~\ref{fig:pptoA} for the LHC runs at $14\,\TeV$, where the CP-odd Higgs boson is considered in the mass range of $M_A\in (300\,\GeV\,, 1\,\TeV)$. We choose the inputs of  $t_\beta=(1\,,5\,,10)$ for the 2HDM-I case and $t_\beta=(1\,,5\,,20)$ for the 2HDM-II case respectively. It is apparent that the inclusive production cross sections of $\sigma[pp\to AX]$ can become sizable with the large $t_\beta$ inputs for the 2HDM-II case, where the bottom quark associated processes become significant. For example, unlike in the 2HDM-I case where the inclusive production cross section of the Higgs boson $A$ decreases with increasing $t_\beta$, the production cross section increases in the 2HDM-II case when the $t_\beta$ value is increased from $t_\beta=5$ to $t_\beta=20$, as shown in plot-(b) of Fig.~\ref{fig:pptoA}.

\subsection{The decay modes and search signals of $A$}

The tree-level decay channels of $A$ in our discussions here include: $A\to (\bar f f\,,hZ\,,HZ)$, with $f$ being the SM fermions. These partial decay widths are expressed as
\beqs
\beqn
\Gamma[A\to \bar f f]&=&\frac{N_{c,f} m_{f}^{2} M_{A}}{8\pi v^{2}}(\xi_{A}^{f})^{2}\sqrt{1-\frac{4m_{f}^{2}}{M_{A}^{2}}}\,,\label{eq:AffWid}\\
\Gamma[A\to hZ]&=&\frac{g^{2}c_{\beta-\alpha}^{2}}{64\pi M_{A}c_{W}^{2}}\lambda^{1/2} \Big( 1\,,\frac{m_{Z}^{2}}{M_{A}^{2}}\,,\frac{M_{h}^{2}}{M_{A}^{2}} \Big)\non
&\times&\Big[ m_{Z}^{2}-2(M_{A}^{2}+M_{h}^{2})+\frac{(M_{A}^{2}-M_{h}^{2})^{2}}{m_{Z}^{2}}  \Big]\,,\label{eq:AhZWid}\\
\Gamma[A\to HZ]&=&\frac{g^{2}s_{\beta-\alpha}^{2}}{64\pi M_{A}c_{W}^{2}}\lambda^{1/2} \Big( 1\,,\frac{m_{Z}^{2}}{M_{A}^{2}}\,,\frac{M_H^{2}}{M_{A}^{2}} \Big)\non
&\times&\Big[ m_{Z}^{2}-2(M_{A}^{2}+M_H^{2})+\frac{(M_{A}^{2}-M_H^{2})^{2}}{m_{Z}^{2}}  \Big]\,,\label{eq:AHZWid}
\eeqn
\eeqs
with $N_{c, f}=3\,(1)$ for quarks (leptons). The three-body phase space factor reads
\beqn\label{eq:threebody_phase}
\lambda^{1/2}(1\,,x^2 \,, y^2)&\equiv& \Big[ ( 1-x^2-y^2 )^2 - 4x^2 y^2  \Big]^{1/2}\,.
\eeqn
For the $M_h = M_H = 125\,\GeV$ degenerate scenario, where one cannot discriminate between $A\to hZ$ and $A\to HZ$, one should add up these two decay channels, $\Gamma[A\to hZ]+ \Gamma[A\to HZ]$, which is collectively denoted as $\Gamma[A\to hZ]$ again in this special case. Thus, the partial width of $\Gamma[A\to hZ]$ in the degenerate scenario becomes independent of the alignment parameter $c_{\beta - \alpha}$
\beqn\label{eq:AhHZWid}
\Gamma[A\to hZ]_{\rm deg}&=&\Gamma[A\to hZ]+\Gamma[A\to HZ]\non
&=&\frac{g^{2} }{64\pi M_{A}c_{W}^{2}}\lambda^{1/2} \Big( 1\,,\frac{m_{Z}^{2}}{M_{A}^{2}}\,,\frac{M_{h}^{2}}{M_{A}^{2}} \Big)\non
&\times&\Big[ m_{Z}^{2}-2(M_{A}^{2}+M_{h}^{2})+\frac{(M_{A}^{2}-M_{h}^{2})^{2}}{m_{Z}^{2}}  \Big]\,.
\eeqn
The loop-induced partial decay width of $\Gamma[A\to gg]$ was given in Eq.~(\ref{eq:Atogg}), while other decay widths of $\Gamma[A\to \gamma\gamma]$ and $\Gamma[A\to Z\gamma]$ are typically negligible.

\begin{figure}
\centering
\includegraphics[width=6.8cm,height=4cm]{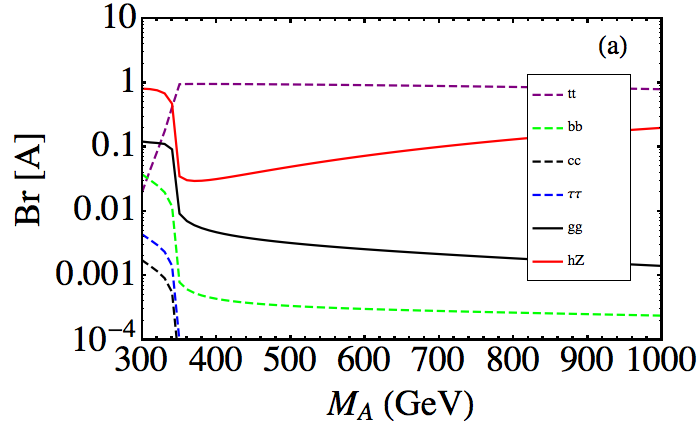}
\includegraphics[width=6.8cm,height=4cm]{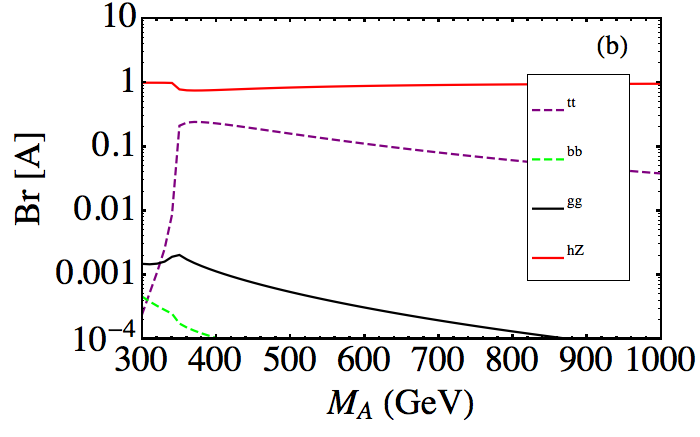}\\
\includegraphics[width=6.8cm,height=4cm]{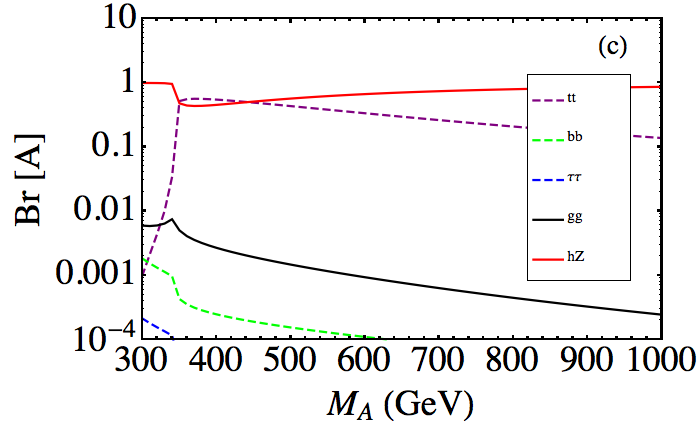}
\includegraphics[width=6.8cm,height=4cm]{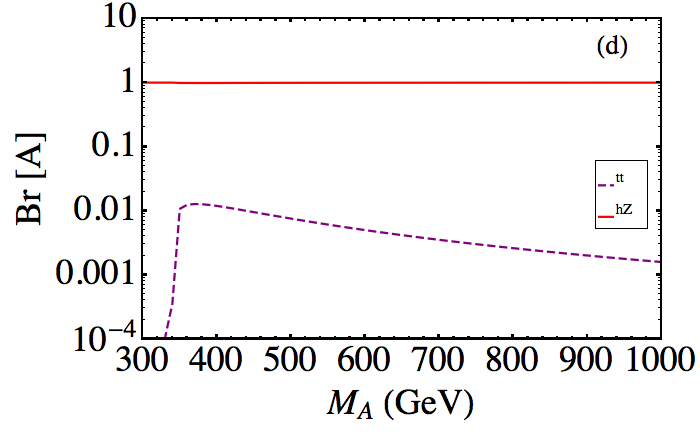}
\caption{\label{fig:BRAI} The decay branching ratios of the CP-odd Higgs boson ${\rm BR}[A]$ for the 2HDM-I case. Upper left: $M_h=125\,\GeV$ with $t_\beta=1$. Upper right: $M_h=125\,\GeV$ with $t_\beta=10$. Lower left: $M_h=M_H=125\,\GeV$ with $t_\beta=1$. Lower right: $M_h=M_H=125\,\GeV$ with $t_\beta=10$. The decay channels with branching ratios below $10^{-4}$ are not shown. }
\end{figure}

\begin{figure}
\centering
\includegraphics[width=6.8cm,height=4cm]{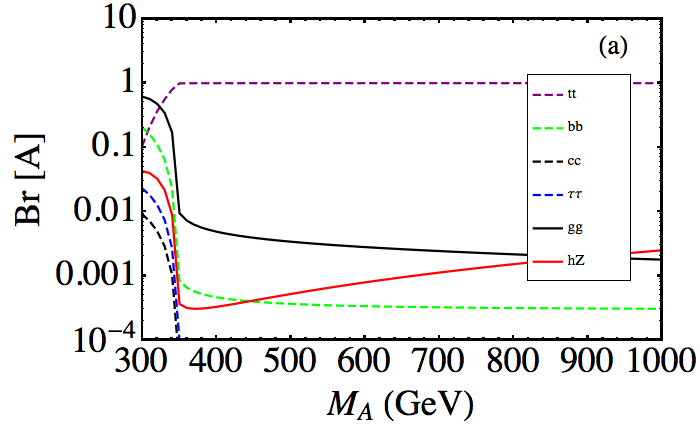}
\includegraphics[width=6.8cm,height=4cm]{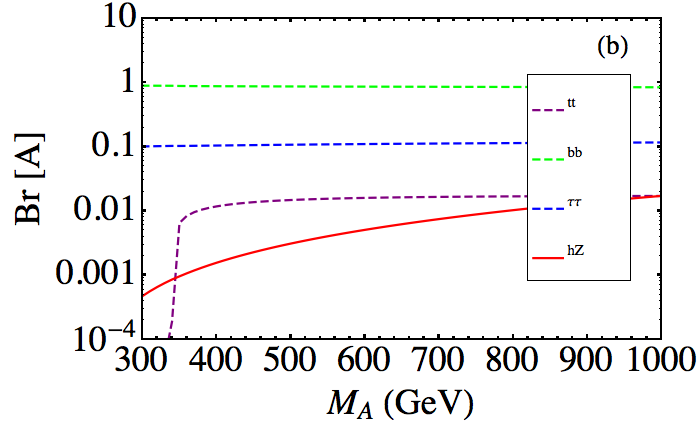}\\
\includegraphics[width=6.8cm,height=4cm]{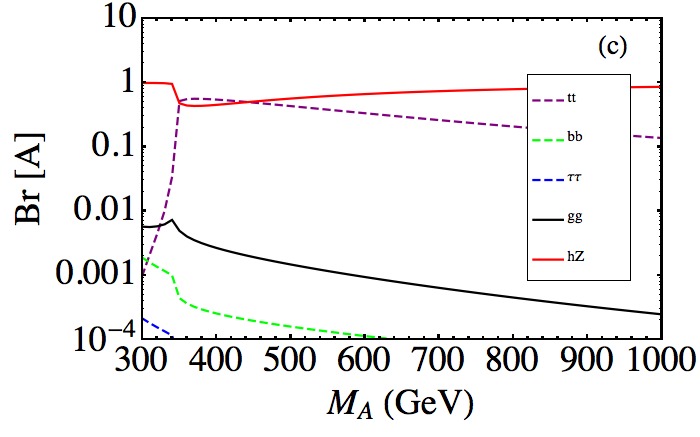}
\includegraphics[width=6.8cm,height=4cm]{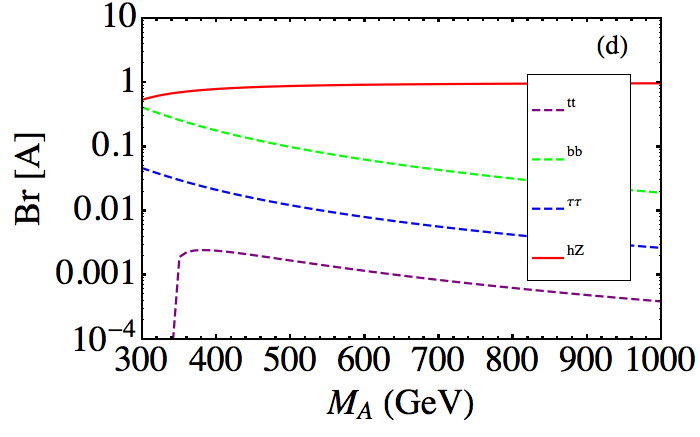}
\caption{\label{fig:BRAII} The decay branching ratios of the CP-odd Higgs boson ${\rm BR}[A]$ for the 2HDM-II case. Upper left: $M_h=125\,\GeV$ with $t_\beta=1$. Upper right: $M_h=125\,\GeV$ with $t_\beta=20$. Lower left: $M_h=M_H=125\,\GeV$ with $t_\beta=1$. Lower right: $M_h=M_H=125\,\GeV$ with $t_\beta=20$. The decay channels with branching ratios below $10^{-4}$ are not shown.}
\end{figure}

\begin{figure}
\centering
\includegraphics[width=6.8cm,height=4cm]{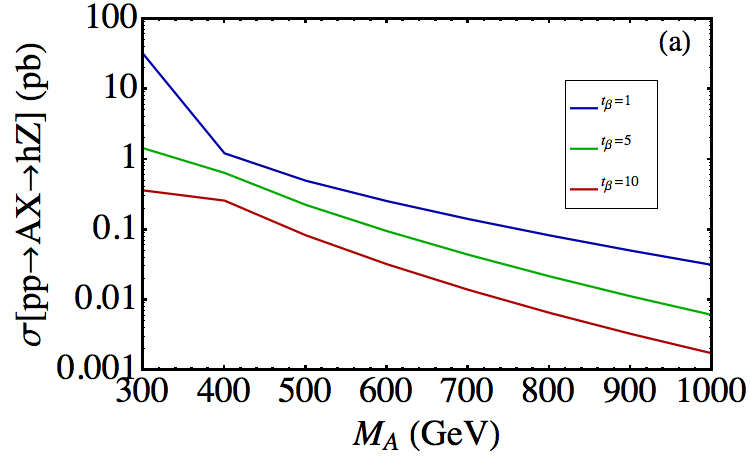}
\includegraphics[width=6.8cm,height=4cm]{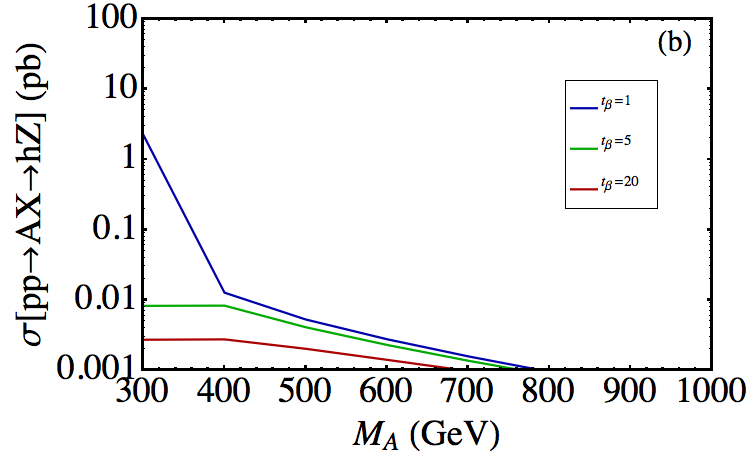}\\
\includegraphics[width=6.8cm,height=4cm]{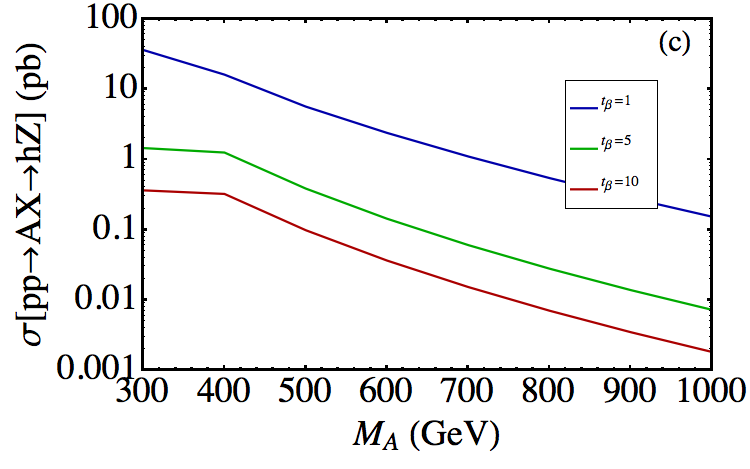}
\includegraphics[width=6.8cm,height=4cm]{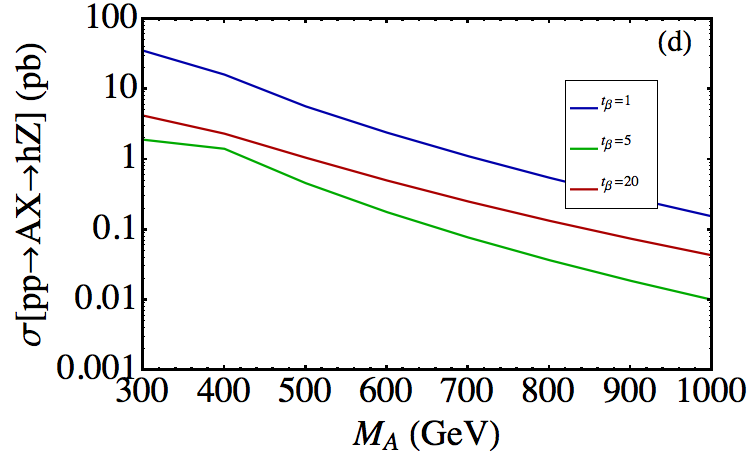}
\caption{\label{fig:xsecA} The $\sigma[pp\to AX]\times {\rm BR}[A\to hZ]$ for $M_A\in (300\,\GeV\,, 1\,\TeV)$ at the LHC $14\,\TeV$ runs. Upper left: $M_h=125\,\GeV$ for 2HDM-I. Upper right: $M_h=125\,\GeV$ for 2HDM-II. Lower left: $M_h=M_H=125\,\GeV$ for 2HDM-I. Lower right: $M_h=M_H=125\,\GeV$ for 2HDM-II. }
\end{figure}

Among all fermionic decay modes, the $A\to \bar t  t$ is generally the most dominant one except for the large $t_\beta$ regions in the 2HDM-II case. It is interesting to compare the partial decay widths of $\Gamma[A\to \bar t t]$ and $\Gamma[A\to hZ]$ in the $M_A\gg (m_Z\,,M_h)$ limit
\beqn\label{eq:AhZtoAtt}
\frac{\Gamma[A\to \bar t t]}{\Gamma[A\to hZ]}&\approx&\frac{8 N_{c\,,f} m_t^2 c_W^2 m_Z^2}{g^2 v^2 M_A^2 t_\beta^2 c_{\beta-\alpha}^2}= 6\Big( \frac{m_t}{M_A} \Big)^2 \frac{1}{ t_\beta^2  c_{\beta-\alpha}^2 }\,.
\eeqn
With the large CP-odd Higgs boson mass of $M_A\gtrsim 2m_t$, it is quite possible to have $\Gamma[A\to \bar t t]\ll \Gamma[A\to hZ]$ with the 2HDM parameters being $c_{\beta-\alpha}^2 t_\beta^2 \gtrsim \mO(1)$. Further considering the degenerate scenario of $M_h=M_H=125\,\GeV$, the alignment parameter does not enter Eq.~(\ref{eq:AhHZWid}). Correspondingly, the decay mode of $A\to hZ$ would dominate over the $A\to \bar t t$ mode with $M_A \gtrsim 2m_t$. In Figs.~\ref{fig:BRAI} and \ref{fig:BRAII}, we display the decay branching ratios of the CP-odd Higgs boson $A$ in the mass range of $M_A\in (300\,\GeV\,, 1\,\TeV)$ for the 2HDM-I and 2HDM-II cases respectively. In practice, the decay branching ratios of the CP-odd Higgs boson demonstrated here are evaluated by {\tt 2HDMC-1.6.4}~\cite{Eriksson:2009ws}. In Figs.~\ref{fig:BRAI} and \ref{fig:BRAII}, we demonstrate the branching ratios for both the $M_h = 125\,\GeV$ scenario and the $M_h = M_H = 125 \,\GeV$ degenerate scenario. The decay branching ratios of ${\rm BR}[A\to hZ]$ are increasing with the larger $M_A$ and $t_\beta$ inputs. For the $M_h = 125\,\GeV$ scenario, the ${\rm BR}[A\to hZ]$ increases from $\mO(0.1)$ to almost unity with the increase of $t_\beta$ from $1$ to $10$ in the 2HDM-I case. However, in the 2HDM-II case, this decay mode is always subdominant for both small and large $t_\beta$ inputs, given the small alignment parameter taken in Eq.~(\ref{eq:align_para}). For the $M_h = M_H = 125\,\GeV$ scenario, the ${\rm BR}[A\to hZ]$ can be the most dominant one over the mass range we are interested in.

Fig.~\ref{fig:xsecA} shows the $\sigma[pp\to AX]\times {\rm BR}[A\to hZ]$ at the LHC $14\,\TeV$ runs by combining the inclusive production cross sections and the decay branching ratios. Based on the analysis to be discussed in the next section, the LHC cross section of $\sigma[pp\to AX]\times {\rm BR}[A\to hZ]$ needs to reach $\sim \mO(0.1)\,\pb$ in order to be probed at the $14\,\TeV$ runs with the integrated luminosity of $\sim \mO(100)\,\fb^{-1}$.

For the $M_h = 125 \,\GeV$ scenario, in the 2HDM-I case, this decay mode of $A\to hZ$ can be the possible search channel for the CP-odd Higgs boson as heavy as $\sim 1\,\TeV$ with $t_\beta$ being not too large; in the 2HDM-II case, however, the search potential to the $A\to hZ$ mode is much smaller, because the cross section in this case is typically small. For the $M_h = M_H = 125 \,\GeV$ degenerate scenario,  the search potential to the $A\to hZ$ mode is significantly improved for both the 2HDM-I and the 2HDM-II cases. By simple counting of the $\sigma[pp\to AX]\times {\rm BR}[A\to hZ]$, one can envision this decay mode to be promising for $M_A$ as large as $\mO(1)\,\TeV$ at the LHC $14\,\TeV$ runs with the integrated luminosity up to $\sim \mO(100)- \mO(10^3)\,\fb^{-1}$. In our analysis below, we shall use the $h/H\to \bar b b$ final states in order to tag the fat Higgs jet. For this reason, the cross sections for the signal processes read
\beqs\label{eqs:signal_xsec}
\beqn
M_h=125\,\GeV&:&\sigma[pp\to AZ]\times {\rm BR}[A\to hZ] \times  {\rm BR}[h\to \bar b b ] \,,\label{eq:signal_nondeg_xsec}\\
M_h = M_H = 125\,\GeV&:&\sigma[pp\to AZ]\times \Big( {\rm BR}[A\to hZ] \times  {\rm BR}[h\to \bar b b ]\non
&& + {\rm BR}[A\to HZ] \times  {\rm BR}[H\to \bar b b ]  \Big)\,,\label{eq:signal_deg_xsec}
\eeqn
\eeqs
respectively. In the $M_h=125\,\GeV$ scenario, the current global fit to the 2HDM parameter regions of $(\alpha\,,\beta)$ point to a SM-like Higgs boson $h$. Hence, it is reasonable to take ${\rm BR}[h\to \bar b b]\approx {\rm BR}[h_{\rm SM}\to \bar b b]=0.58$ for our estimation of the signal cross sections below. In the $M_h=M_H=125\,\GeV$ scenario, however, a global fit to the $125\,\GeV$ Higgs is lacking. One can further write the branching ratios in the Eq.~(\ref{eq:signal_deg_xsec}) as
\beqn
&& {\rm BR}[A\to hZ] \times  {\rm BR}[h\to \bar b b ]+ {\rm BR}[A\to HZ] \times  {\rm BR}[H\to \bar b b ] \non
&=&{\rm BR}[A\to hZ]_{\rm deg}\times \Big(  c_{\beta-\alpha}^2{\rm BR}[h\to \bar b b ] + s_{\beta-\alpha}^2 {\rm BR}[H\to \bar b b ] \Big)\,,
\eeqn
where we used the Eqs.~(\ref{eq:AhZWid}), (\ref{eq:AHZWid}), and (\ref{eq:AhHZWid}) in the last line. Instead of constraining the 2HDM parameters for the $M_h = M_H=125\,\GeV$ scenario, here we assume that the branching ratios in the parenthesis reproduce the SM value, i.e., $c_{\beta-\alpha}^2{\rm BR}[h\to \bar b b ] + s_{\beta-\alpha}^2 {\rm BR}[H\to \bar b b ] \approx {\rm BR}[h_{\rm SM}\to \bar b b]=0.58$. This approximation is reasonable if we assume the future LHC searches for the $\bar b b$ final states via the $pp\to Vh(\to \bar b b)$ process are close to the SM Higgs predictions. 

%The corresponding cross sections of the $pp\to Vh(\to \bar b b)$ process in the $M_h = M_H = 125\,\GeV$ degenerate scenario reads
%
%\beqn
%&&\sigma[pp\to Vh]\times {\rm BR}[h\to \bar b b]  + \sigma[pp\to VH]\times {\rm BR}[H\to \bar b b]\non
%&=&\sigma[pp\to Vh_{\rm SM}]\times \Big( c_{\beta-\alpha}^2 {\rm BR}[h\to \bar b b]  +  s_{\beta-\alpha}^2 {\rm BR}[H\to \bar b b]  \Big)\,.
%\eeqn

%###################################################################

\section{The LHC Searches for The Exotic $A\to hZ$ Channel}
\label{section:AhZ}

In this section, we proceed to analyze the LHC searches for the CP-odd Higgs boson $A$ via the decay mode of $A\to hZ$.

\subsection{The SM backgrounds and signal benchmark}

The final states to be searched for are the same as the ones in the SM Higgs boson searches via the $hZ$ associated production channel. Therefore, the dominant irreducible SM backgrounds relevant to our analysis are~\cite{ATLAS:2013079}: $\bar b b \ell^+ \ell^-$, $\bar t t$, $ZZ \to\bar b b \ell^{+}\ell^{-}$, and the $h_{\rm SM} Z\to \bar b b \ell^{+}\ell^{-}$. The cross sections for these processes~\cite{Ahrens:2011px, Cordero:2009kv, Dittmaier:2011ti,Campbell:2011bn} at the LHC $14\,\TeV$ run read
\beqn\label{eq:SM_bkg}
&&\sigma(pp\to \bar t t)\approx 855\,\pb\,,\non
&&\sigma(pp\to b \bar b \ell^+ \ell^-)\approx 82\,\pb \,,\non
&&\sigma (pp\to ZZ \to \bar b b \ell^+ \ell^-)\approx180\,\fb\,,\non
&&\sigma(pp\to h_{\rm SM} Z \to \bar b b \ell^+ \ell^-  )\approx 34\,\fb\,.
\eeqn
In practice, we note the major SM background processes of $\bar t t$ and $\bar b b \ell^+ \ell^-$ receive uncertainties of $\sim 9\%$ and $\sim 14\%$ respectively. In our analysis below, we take the $b-$tagging efficiency of $70\,\%$~\cite{ATLAS:2012aoa}, and the mis-tagging rates are taken as
\beqn\label{eq:bfake}
&&\epsilon_{c\to b}\approx 0.2 \qquad \epsilon_{j\to b}\approx 0.01\,,
\eeqn
with $j$ representing the light jets that neither originate from a $b$ quark nor a $c$ quark~\cite{ATLAS:2012lma}.

In order to generate events for the signal processes, we obtain a Universal FeynRules Output~\cite{Christensen:2008py} simplified model with $A$ being the only BSM particle. The relevant coupling terms are implemented, namely, the dimension-five $Agg$ coupling, the derivative coupling of $AhZ$, and the $A(h)\bar b b$ Yukawa couplings. We generate events at the parton level by {\tt Madgraph 5}~\cite{Alwall:2014hca}, which are passed to {\tt Pythia}~\cite{Sjostrand:2006za} for the parton showering and hadronization. In order to employ the fat Higgs jet tagging method~\cite{Butterworth:2008iy}, the $B$-hadron decays are turned off. All events are further passed to {\tt Delphes-3.1.2}~\cite{deFavereau:2013fsa} for the fast detector simulation, where we apply the default ATLAS detector card.

\subsection{The jet substructure methods}

Here we describe the jet substructure analysis and the application to the signals we are interested in. We pass the events to {\tt Fastjet}~\cite{Cacciari:2011ma} in order to cluster the final states. The tracks, neutral hadrons, and photons that enter the jet reconstruction should satisfy $p_T > 0.1\,\GeV$ and $|\eta|<5.0$. The leptons from the events should be isolated, so that they will not be used to cluster the fat jets. The fat jets are reconstructed by using the C/A jet algorithm with particular jet cone size $R$ to be specified below and requiring $p_T>30\,\GeV$. Afterwards, we adopt the procedures described in the mass-drop tagger~\cite{Butterworth:2008iy} for the purpose of identifying a boosted Higgs boson:

\begin{itemize}

\item Split the fat jet, $j$, into two subjets $j_{1\,,2}$ with masses $m_{1\,,2}$, and $m_1 > m_2$.

\item Require a significant mass drop of $m_1 < \mu m_j$ with $\mu = 0.667$, and also a sufficiently symmetric splitting of ${\rm min} ( p_{T\,,1}^2\,, p_{T\,,2}^2) \Delta R_{12}^2 / m_j^2 > y_{\rm cut}$ ($\Delta R_{12}^2$ is the angular distance between $j_1$ and $j_2$ on the $\eta-\phi$ plane) with $y_{\rm cut}=0.09$.

\item If the above criteria are not satisfied, define $j\equiv j_1$ and go back to the first step for decomposition.

\end{itemize}
These steps are followed by the filtering stage using the reclustering radius of $R_{\rm filt} = {\rm min} (0.35\,, R_{12}/2)$ and selecting three hardest subjects to suppress the pile-up effects. 

\begin{figure}
\centering
\includegraphics[width=3.4cm,height=3cm]{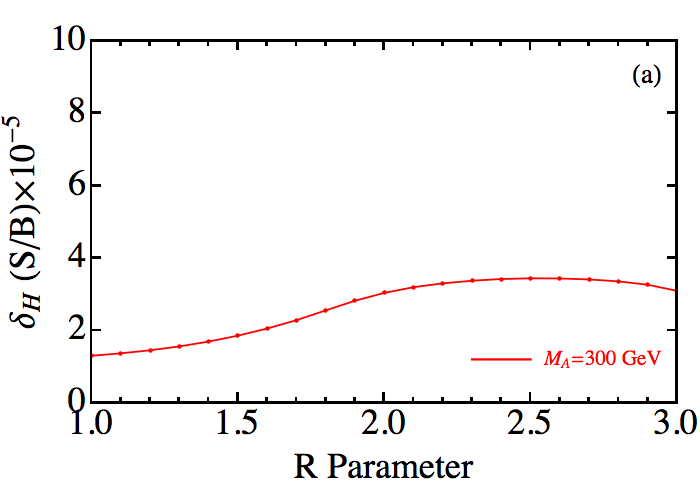}
\includegraphics[width=3.4cm,height=3cm]{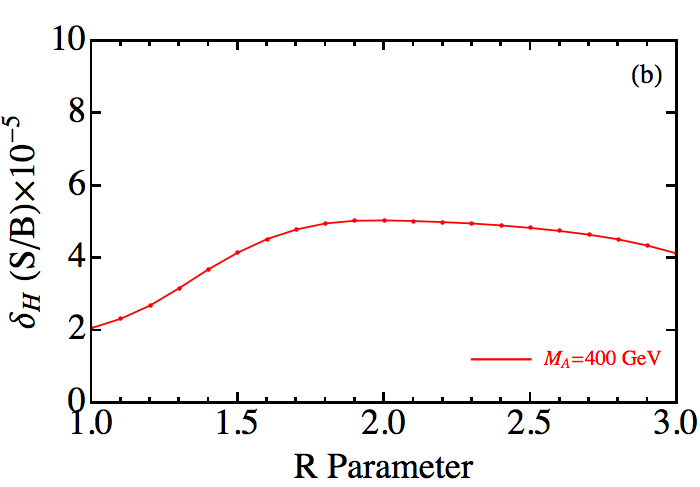}
\includegraphics[width=3.4cm,height=3cm]{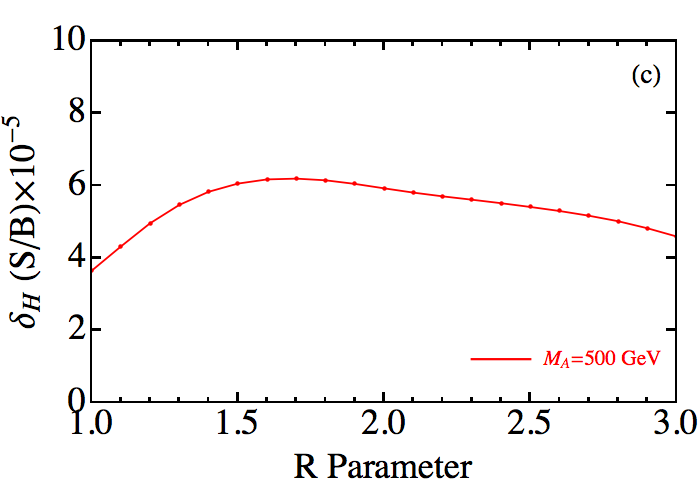}
\includegraphics[width=3.4cm,height=3cm]{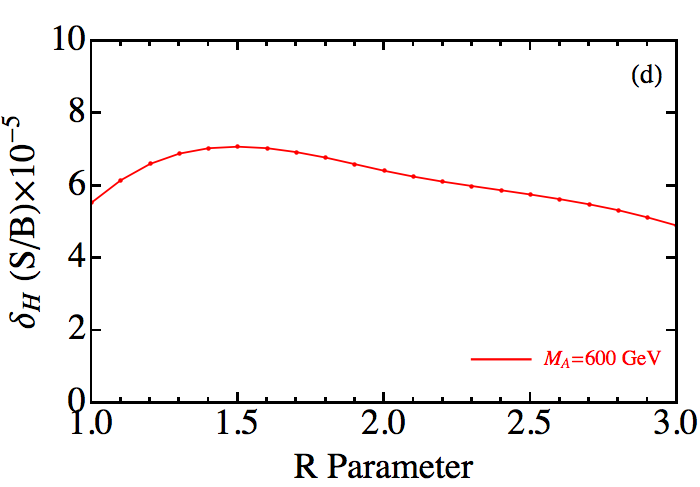}\\
\includegraphics[width=3.4cm,height=3cm]{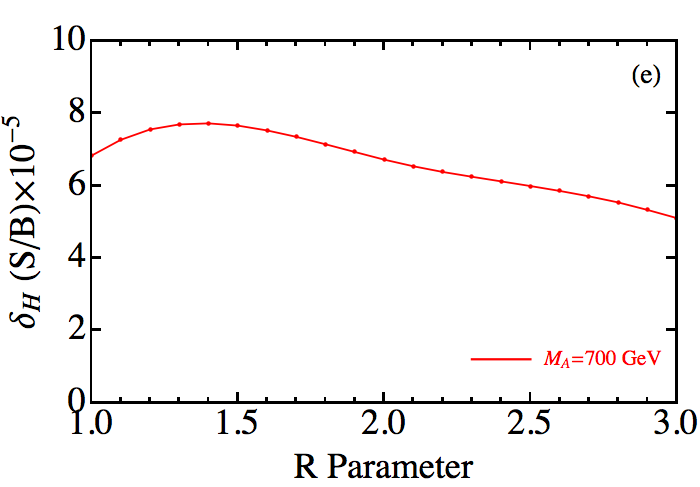}
\includegraphics[width=3.4cm,height=3cm]{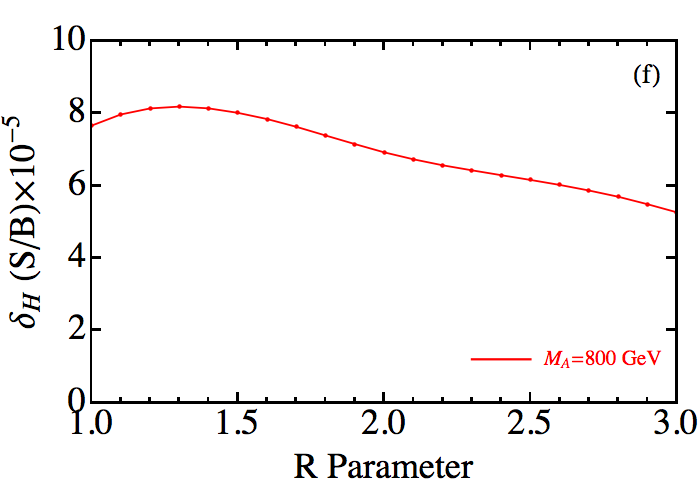}
\includegraphics[width=3.4cm,height=3cm]{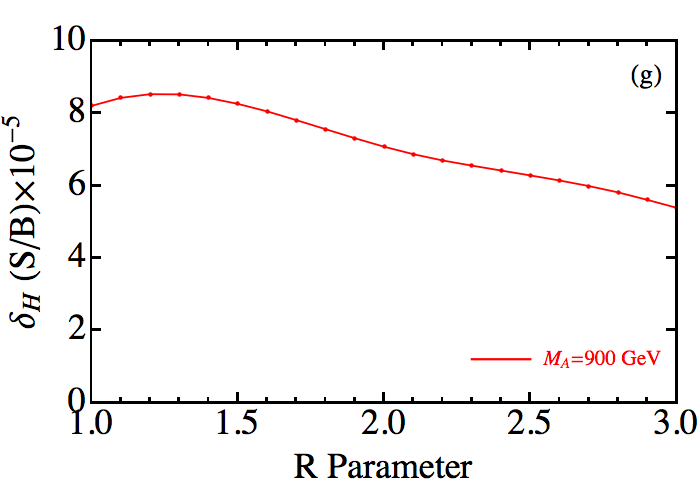}
\includegraphics[width=3.4cm,height=3cm]{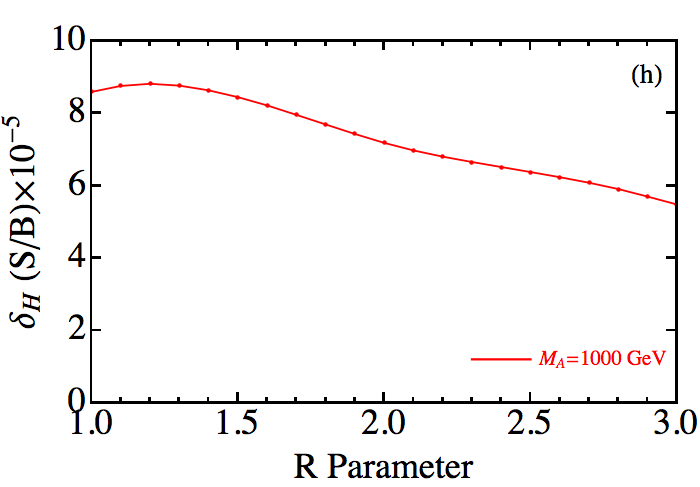}
\caption{\label{fig:Rvalues} The fat Higgs jet tagging rates $\delta_H(S/B)$ with the varying jet cone sizes $R$ in the C/A jet algorithm. For comparison, we take a common cross section of $\sigma[pp\to AX\to hZ]=100\,\fb$ for all signal processes.}
\end{figure}

\begin{table}[htb]
\begin{center}
\begin{tabular}{c|c|c|c|c}
\hline
%\hline
$M_A$ & $300\,\GeV$ & $400\,\GeV$ & $500\,\GeV$  & $600\,\GeV$    \\\hline
 C/A algorithm $R$ & $2.5$ & $2.0$ & $1.7$ & $1.5$ \\\hline\hline
 $M_A$ & $700\,\GeV$ & $800\,\GeV$ & $900\,\GeV$  & $1000\,\GeV$    \\\hline
 C/A algorithm $R$ & $1.4$ & $1.3$ & $1.2$ & $1.2$ \\
 \hline
\end{tabular}
\caption{The choices of the jet cone sizes $R$ in the C/A jet algorithm for different $M_A$ inputs. }\label{tab:Ropt}
\end{center}
\end{table}

Generally, the jet cone size $R$ taken in the C/A algorithm tends to be large in order to capture all collimated decay products in a fat jet. Since our final states involve a SM-like Higgs boson $h$ from the $A\to hZ$ decay, the corresponding boost factors are enhanced for the larger $M_A$ case. To determine the most optimal jet cone size $R$ in the C/A jet algorithm choice for each $M_A$ input, we vary it in the range of $1.0\leq R\leq 3.0$ and look for the maximal fat Higgs jet tagging rates $\delta_H(S/B)$ 
\beqn
\delta_H (S/B)&\equiv&\frac{\textrm{ number of Higgs jets tagged in the signal }}{\sum_{\rm background} \textrm{number of Higgs jets tagged in SM background}}
\eeqn
between the signals and SM backgrounds. In Fig.~\ref{fig:Rvalues}, we demonstrated the fat Higgs jet tagging rate $\delta_H$ for different $M_A$ samples with the varying $1.0\leq R\leq 3.0$. Accordingly, the most optimal jet cone size $R$ to be chosen for each $M_A$ input is tabulated in Table.~\ref{tab:Ropt}. As seen from the table, a smaller cone size $R$ is generally favored for the heavier CP-odd Higgs boson.

\subsection{The event selection}

The cut flow we impose to the events are the following:

\begin{itemize}

\item Cut 1: We select events with the opposite-sign-same-flavor (OSSF) dileptons $(\ell^+ \ell^-)$ in order to reconstruct the final-state $Z$ boson. The OSSF dileptons are required to satisfy the following selection cuts

\beqn
&&|\eta_\ell|<2.5\,,~~~ p_T(\ell_1)\ge 20\,\GeV\,,~~~ p_T(\ell_2)\ge 10\,\GeV\,,
\eeqn
where $\ell_{1\,,2}$ represent two leading leptons ordered by their transverse momenta.

\item Cut 2: The invariant mass of the selected OSSF dileptons should be around the mass window of $Z$ boson $|m_{\ell\ell} - m_Z|\le 15\,\GeV$.

\item Cut 3: At least one filtered fat jet is required, which should also contain two leading subjets that pass the b-tagging and satisfy $p_T> 20\,\GeV$ and $|\eta|<2.5$.

\item Cut 4: Such a filtered fat jet will be then identified as the SM-like Higgs jet. We impose the cuts to the filtered Higgs jets in the mass window of $M_h({\rm tagged})\in (100\,\GeV\,,150\,\GeV)$.

\begin{figure}
\centering
\includegraphics[width=6.8cm,height=4.2cm]{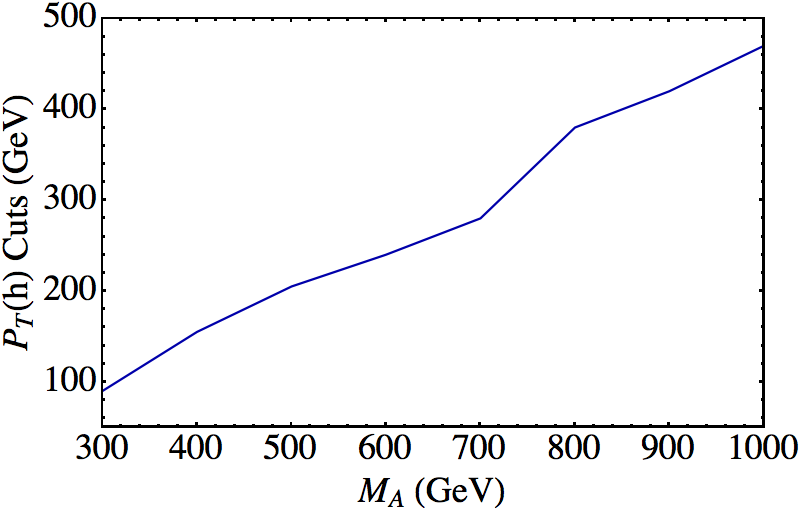}
\caption{\label{fig:PTopt} The most optimal cuts to the $p_T$ of the tagged SM-like Higgs boson for different $M_A$ inputs.}
\end{figure}

\item Cut 5: We also impose the cuts on the $p_{T\,,h}({\rm tagged})$. The SM-like Higgs bosons decaying from the heavier CP-odd Higgs $A$ would generally be more boosted. In practice, we vary the $p_{T\,,h}({\rm tagged})_{\rm cut}\in (50\,\GeV\,,500\,\GeV)$ and look for the most optimal cuts on $p_{T\,,h}({\rm tagged})$ by counting the corresponding cut efficiencies of $S/B$. The $p_{T\,,h}({\rm tagged})$ cuts to be adopted below are displayed in Fig.~(\ref{fig:PTopt}).

\item Cut 6: Combining the filtered Higgs jets and the tagged OSSF dileptons, the invariant mass of the tagged Higgs boson and the OSSF leptons should reconstruct the mass window of the CP-odd Higgs boson $A$: $|M_{h\,,\ell^+\ell^-} - M_A|\leq 100\,\GeV$.

\end{itemize}

\subsection{Implications to the LHC searches for $A$ in the general 2HDM}

%%%%%%%%%%%%%%%%%%%%%%%%%%%%%%%%%%%%%%%%%%%%%%%%%%%%%%%
\begin{table}[t]
\begin{center}
\begin{tabular}{c|c|cccccc}
\hline
%\hline
 Cuts & $A\to hZ$ & $\bar t t$ & $\bar b b \ell^+ \ell^-$ & $ZZ\to \bar b b\ell^+\ell^-$ & $hZ\to \bar b b\ell^+\ell^-$ & $S/B$ & $S/\sqrt{B}$    \\\hline\hline  
 Total cross section $(\fb)$ & $500$ & $8.6\times 10^5$ & $8.2\times 10^4$ & $180$ & $34$ & $-$ &  $$  \\
 Cut 1 & $10.76$  & $1.0\times 10^4$ & $4.3\times 10^4$ & $98.94$ & $0.81$ & $1.3\times 10^{-4}$ & $0.47$ \\
 Cut 2 & $10.29$  & $2,061$ & $3.9\times 10^4$ & $93.49$ & $0.78$ & $1.6\times 10^{-4}$ & $0.51$ \\
 Cut 3 & $2.41$  & $120.63$ & $1,759$ & $4.92$ & $0.05$ & $8.2\times 10^{-4}$ & $0.56$ \\
 Cut 4 & $1.38$  & $13.12$ & $100.54$ & $1.12$ & $0.03$ & $7.7\times 10^{-3}$ & $1.29$ \\
 Cut 5 & $0.91$  & $0.38$ & $12.14$ & $0.19$ & $0.01$ & $0.04$ & $2.55$ \\
 Cut 6 & $0.91$  & $0.06$ & $5.40$ & $0.08$ & $-$ & $0.10$ & $3.87$ \\
\hline
 \end{tabular}
\caption{The event cut efficiency for the $M_A=600\,\GeV$ case at the LHC $14\,\TeV$ running of the signal and background processes. We assume the nominal cross section for the signal process to be $\sigma[pp\to AX]\times {\rm BR}[A\to hZ]=500\,\fb$. The $S/\sqrt{B}$ is evaluated for the $\int \mL dt=100\,\fb^{-1}$ case. The uncertainties of the SM background processes are taken into account.}\label{tab:MA600eff_14TeV}
\end{center}
\end{table}
%%%%%%%%%%%%%%%%%%%%%%%%%%%%%%%%%%%%%%%%%%%%%%%%%%%%%%%

\begin{figure}
\centering
\includegraphics[width=8cm,height=4.5cm]{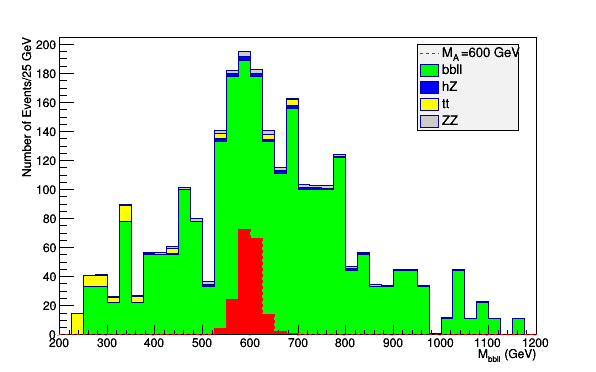}
\caption{\label{fig:AhZ600SB} The $M_{h\,,ll}$ distributions of the $pp \to AX\to hZ$ signal process (for the $M_A=600\,\GeV$ case) and all SM background processes after the kinematic cuts. A nominal cross section of $\sigma[pp\to AX]\times {\rm BR}[A\to hZ]=500\,\fb$ is assumed for the signal. The plot is for the LHC $14\,\TeV$ run with integrated luminosity of $\int\mL dt=100\,\fb^{-1}$. }
\end{figure}

Here we present the results after the jet substructure analysis and imposing the kinematic cuts stated previously. As a specific example of the analysis stated above, we list the cut efficiencies for the benchmark model for the $M_A= 600\,\GeV$ case in Table.~\ref{tab:MA600eff_14TeV}. The distributions of the $M_{h\,,\ell\ell}$ after Cut-1 through Cut-5 for both signal process and the relevant SM background processes are displayed in Fig.~\ref{fig:AhZ600SB}. A nominal production cross section of $\sigma[pp \to AX]\times {\rm BR}[A\to hZ]=500\,\fb$ for the signal process is chosen for the evaluation. Among all relevant SM background processes, the $\bar b b \ell^+ \ell^-$ turns out to contribute most after imposing the cuts mentioned above.

\begin{figure}
\centering
\includegraphics[width=6cm,height=3.5cm]{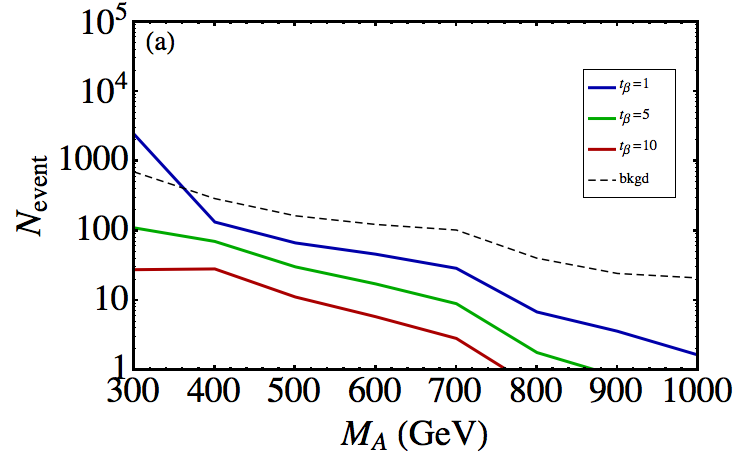}
\includegraphics[width=6cm,height=3.5cm]{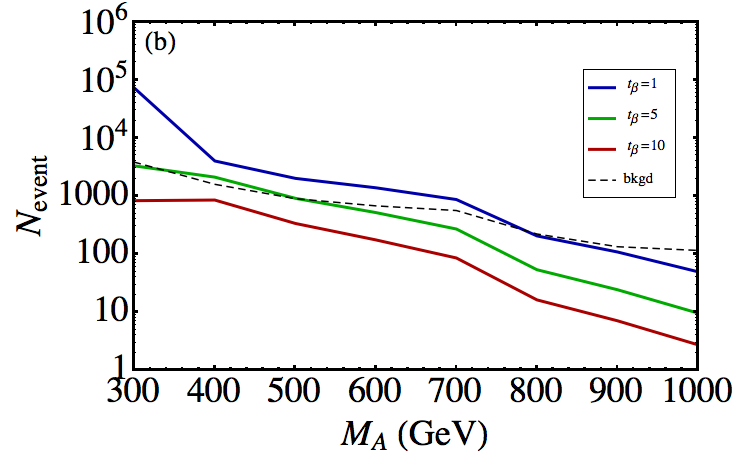}\\
\includegraphics[width=6cm,height=3.5cm]{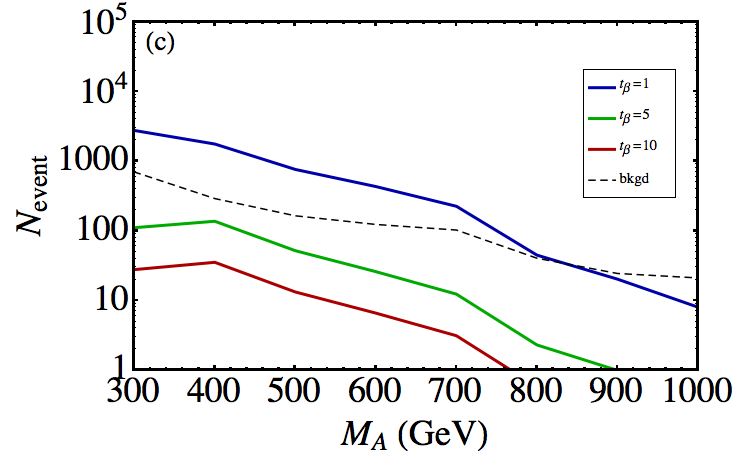}
\includegraphics[width=6cm,height=3.5cm]{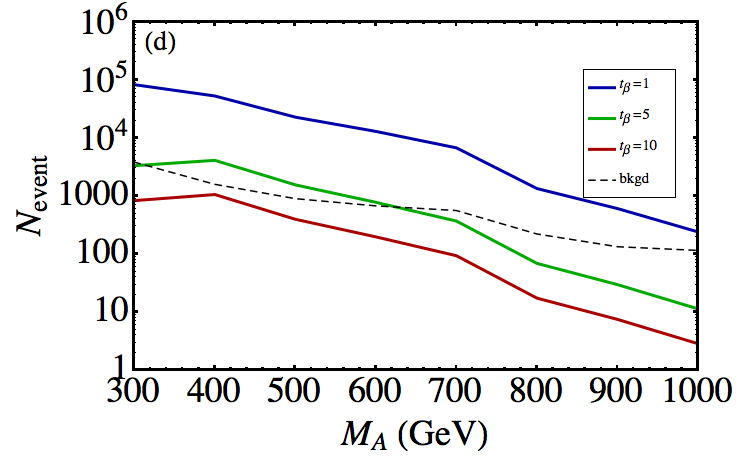}
\caption{\label{fig:AhZIevents} The number of events for the $pp \to AX\to hZ$ signal in the 2HDM-I and the corresponding SM background processes after the jet substructure analysis. Upper left: $M_h=125\,\GeV$ for $\int \mL dt = 100\,\fb^{-1}$. Upper right: $M_h=125\,\GeV$ for $\int \mL dt = 3,000\,\fb^{-1}$. Lower left: $M_h=M_H=125\,\GeV$ for $\int \mL dt = 100\,\fb^{-1}$. Lower right: $M_h=M_H=125\,\GeV$ for $\int \mL dt = 3,000\,\fb^{-1}$. We show samples with $t_\beta=1$ (blue), $t_\beta=5$ (green), and $t_\beta=10$ (red) for each plot. The discovery limit (black dashed curve) of max$\{ 5\sqrt{B}\,,10  \}$ is demonstrated for each plot.}
\end{figure}

\begin{figure}
\centering
\includegraphics[width=6cm,height=3.5cm]{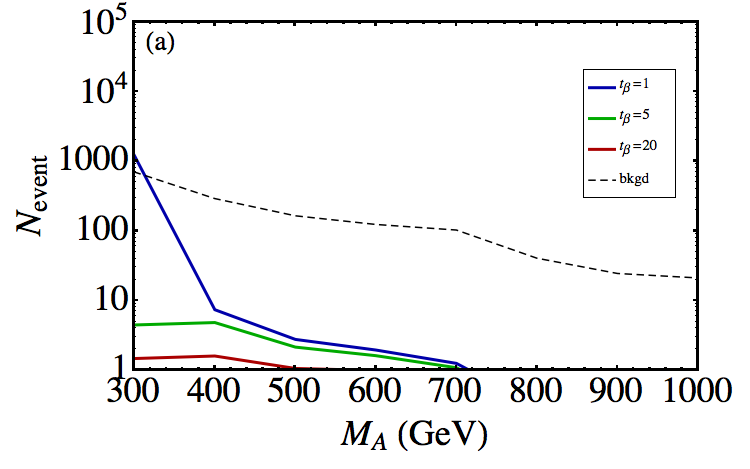}
\includegraphics[width=6cm,height=3.5cm]{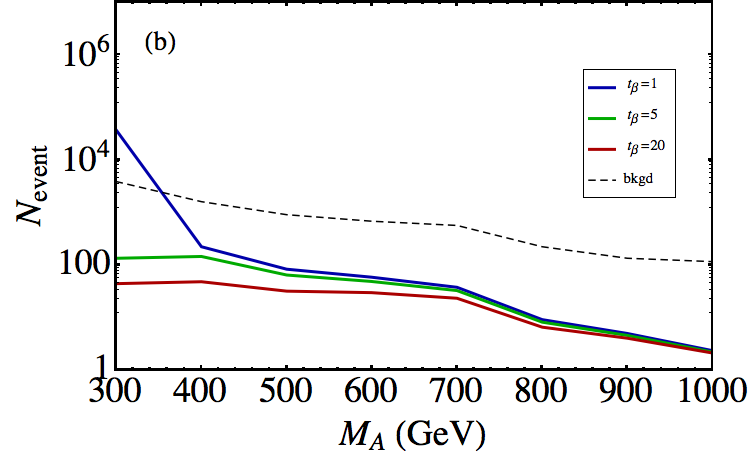}\\
\includegraphics[width=6cm,height=3.5cm]{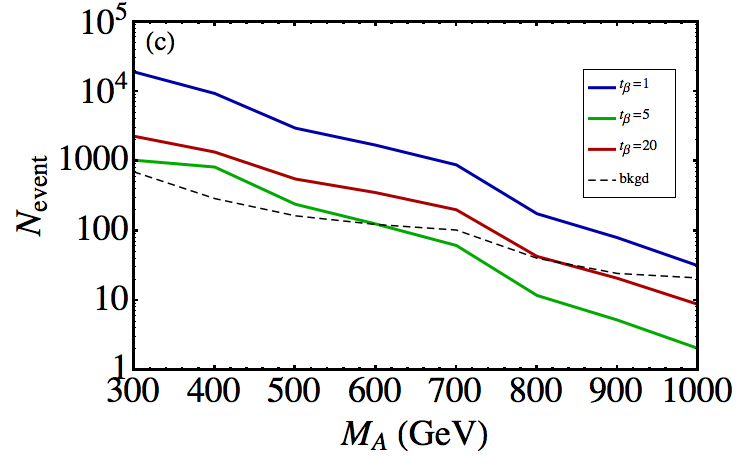}
\includegraphics[width=6cm,height=3.5cm]{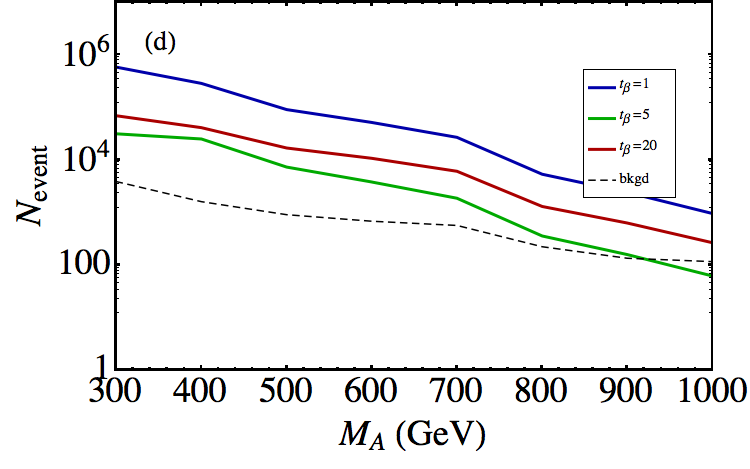}
\caption{\label{fig:AhZIIevents} The number of events for the $pp \to AX\to hZ$ signal in the 2HDM-II and the corresponding SM background processes after the jet substructure analysis. Upper left: $M_h=125\,\GeV$ for $\int \mL dt = 100\,\fb^{-1}$. Upper right: $M_h=125\,\GeV$ for $\int \mL dt = 3,000\,\fb^{-1}$. Lower left: $M_h=M_H=125\,\GeV$ for $\int \mL dt = 100\,\fb^{-1}$. Lower right: $M_h=M_H=125\,\GeV$ for $\int \mL dt = 3,000\,\fb^{-1}$. We show samples with $t_\beta=1$ (blue), $t_\beta=5$ (green), and $t_\beta=10$ (red) for each plot. The discovery limit (black dashed curve) of max$\{ 5\sqrt{B}\,,10  \}$ is demonstrated for each plot.}
\end{figure}

In Figs.~\ref{fig:AhZIevents} and \ref{fig:AhZIIevents}, we display the number of events predicted by the signal process of $pp\to AX\to hZ$ after the cut flows imposed to the 2HDM-I and 2HDM-II models respectively. For each $M_A$ sample, the same kinematic cuts were also imposed to the SM background processes. The samples with different $t_\beta$ inputs are shown for both $M_h = 125\,\GeV$ scenario and $ M_h = M_H = 125\,\GeV $ degenerate scenario. We demonstrate the predictions at the LHC $14\,\TeV$ runs with integrated luminosities of $100\,\fb^{-1}$ and high luminosity (HL) runs up to $3,000\,\fb^{-1}$. Altogether, the $5\sigma$ discovery limits set by max$\{ 5\sqrt{B}\,,10\}$ with $B$ representing the number of events from the SM background contributions are also shown. For the 2HDM-I cases, the $M_h = 125\,\GeV$ scenario consistent to the current global fit to the 2HDM parameter is likely to be probed with $M_A$ up to $1\,\TeV$ with the integrated luminosity $\sim 3,000\,\fb^{-1}$. For the special $M_h = M_H = 125\,\GeV$ degenerate scenario, the discovery limit to the $M_A$ can reach $\sim 1\,\TeV$ at the LHC $14\,\TeV$ runs with $\int\mL dt \sim 100\,\fb^{-1}$. The increasing integrated luminosities would further enhance the discovery limits for models with larger $t_\beta$ inputs. Situations for the 2HDM-II cases are different. The $M_h = 125\,\GeV$ scenario is not promising even at the HL LHC runs with integrated luminosities up to $\sim 3,000\,\fb^{-1}$. Only the CP-odd Higgs boson with mass of $M_A\lesssim 2m_t$ is likely to be searched, together within the low-$t_\beta$ regions. On the other hand, the $M_h = M_H = 125\,\GeV$ degenerate scenario is promising to search for, as indicated from the previous results shown in plot-(d) of Fig.~\ref{fig:xsecA}. As the production cross sections are dominated by the gluon fusion at the low-$t_\beta$ regions, while the bottom quark associated processes can be enhanced at the high-$t_\beta$ regions, the plot-(c) and plot-(d) in Fig.~\ref{fig:AhZIIevents} suggest this channel is promising for the 2HDM-II under the degenerate scenario.

\begin{figure}
\centering
\includegraphics[width=6cm,height=3.5cm]{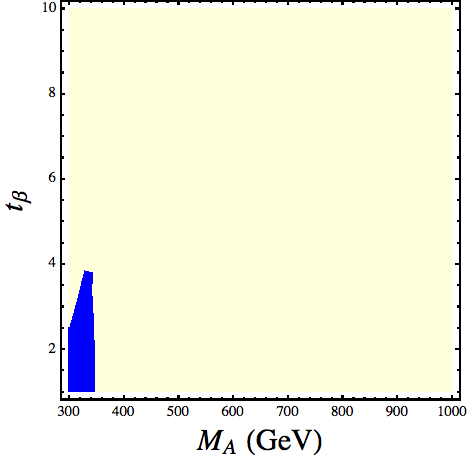}
\includegraphics[width=6cm,height=3.5cm]{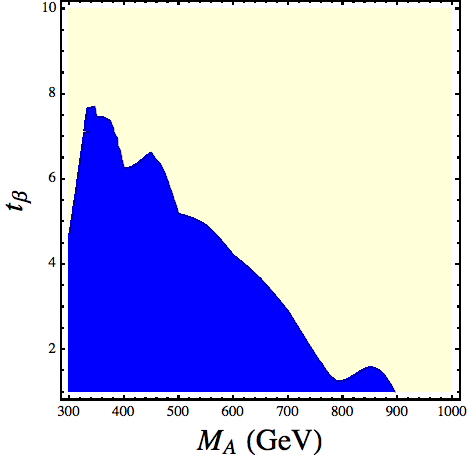}\\
\includegraphics[width=6cm,height=3.5cm]{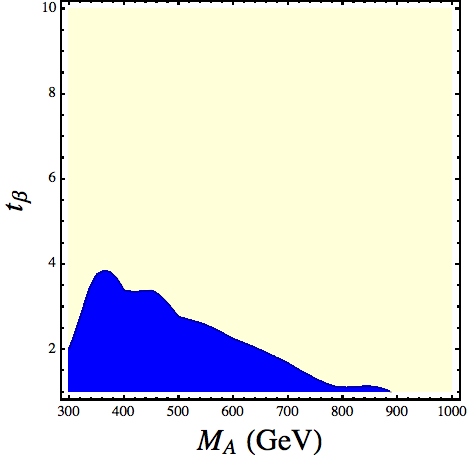}
\includegraphics[width=6cm,height=3.5cm]{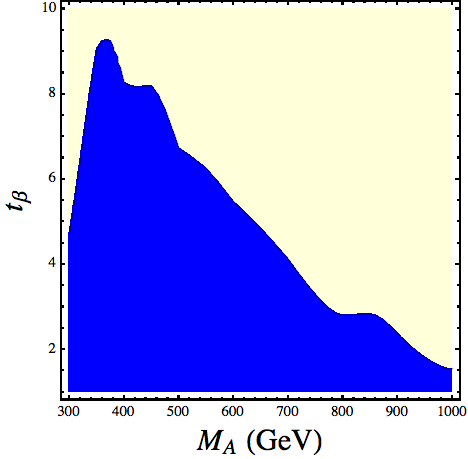}
\caption{\label{fig:AhZIreach} The signal reaches for the $A\to hZ$ on the $(M_A\,, t_\beta)$ plane for the 2HDM-I case. Upper left: $M_h=125\,\GeV$ for $\int \mL dt = 100\,\fb^{-1}$. Upper right: $M_h=125\,\GeV$ for $\int \mL dt = 3,000\,\fb^{-1}$. Lower left: $M_h=M_H=125\,\GeV$ for $\int \mL dt = 100\,\fb^{-1}$. Lower right: $M_h=M_H=125\,\GeV$ for $\int \mL dt = 3,000\,\fb^{-1}$. Parameter regions of $(M_A\,, t_\beta)$ in blue are within the reach for each case.}
\end{figure}

\begin{figure}
\centering
\includegraphics[width=6cm,height=3.5cm]{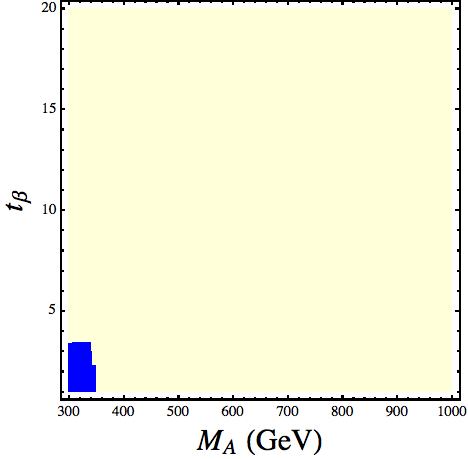}
\includegraphics[width=6cm,height=3.5cm]{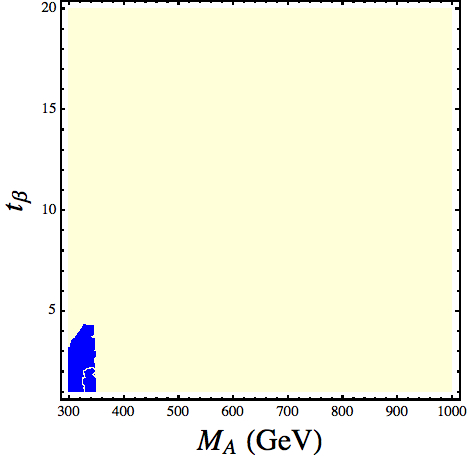}\\
\includegraphics[width=6cm,height=3.5cm]{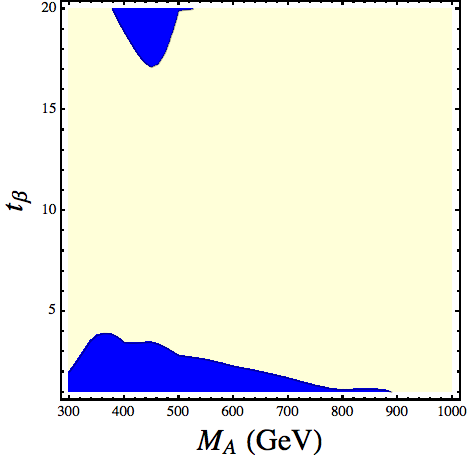}
\includegraphics[width=6cm,height=3.5cm]{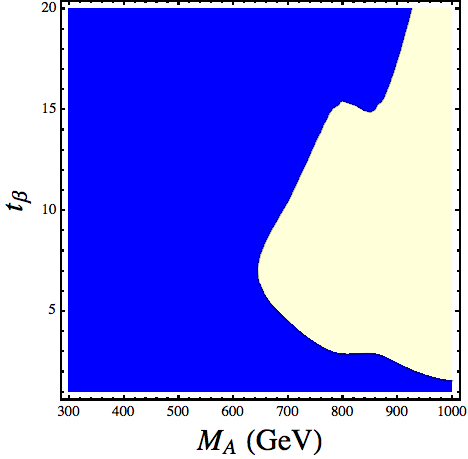}
\caption{\label{fig:AhZIIreach} The signal reaches for the $A\to hZ$ on the $(M_A\,, t_\beta)$ plane for the 2HDM-II case. Upper left: $M_h=125\,\GeV$ for $\int \mL dt = 100\,\fb^{-1}$. Upper right: $M_h=125\,\GeV$ for $\int \mL dt = 3,000\,\fb^{-1}$. Lower left: $M_h=M_H=125\,\GeV$ for $\int \mL dt = 100\,\fb^{-1}$. Lower right: $M_h=M_H=125\,\GeV$ for $\int \mL dt = 3,000\,\fb^{-1}$. Parameter regions of $(M_A\,, t_\beta)$ in blue are within the reach for each case.}
\end{figure}

The signal reaches on the $(M_A\,, t_\beta)$ plane are further displayed in Figs.~\ref{fig:AhZIreach} and \ref{fig:AhZIIreach} for the 2HDM-I and 2HDM-II cases respectively. For the samples we study, both scenarios of $M_h = 125\,\GeV$ and $M_h = M_H = 125\,\GeV$ are shown. There are significant improvements of the signal reaches when increasing the integrated luminosity from $100\,\fb^{-1}$ up to the HL LHC runs up to $3,000\,\fb^{-1}$. For the 2HDM-I case, the $\sigma[pp\to AX]\times {\rm BR}[A\to hZ]$ decreases with the larger $t_\beta$ inputs, as consistent to the plot-(a) and plot-(c) presented in the Fig.~\ref{fig:xsecA}. Correspondingly, this search channel of $A\to hZ$ is generally promising for the low-$t_\beta$ regions. However, for the 2HDM-II case, the large-$t_\beta$ regions are also possible for the search channel of $A\to hZ$. This is true for the special $M_h = M_H = 125\,\GeV$ degenerate scenario. Therefore, one would envision the results presented here are generally complementary to the conventional experimental searches via the $A\to \bar b b$ and $A\to \tau^+ \tau^-$ final states.

%###################################################################

\section{Conclusion}
\label{section:conclusion}

In this work, we suggested that searches for the $hZ$ final states of a heavy CP-odd Higgs $A$ in the general 2HDM can be considered as a potentially promising channel for the upcoming LHC runs at $14\,\TeV$. Such decay channel is due to the derivative coupling term $AhZ$ arising from the 2HDM kinematic terms. Within the framework of the general 2HDM, we consider this decay channel for two scenarios, i.e., the $M_h= 125\,\GeV$ case and the $M_h= M_H= 125\,\GeV$ degenerate Higgs case. For the first scenario, the global fit to the $125\,\GeV$ Higgs boson in the context of the 2HDM is applied. By comparing the decay branching ratios of ${\rm BR}[A\to h Z]$ with other decay modes, together with the evaluation of the inclusive production cross sections for the CP-odd Higgs boson, it is shown that this channel can become the leading one for consideration. Furthermore, the technique of tagging the boosted Higgs jets from the $A\to hZ$ decay chain is very efficient for suppressing the SM background contributions. We optimized the jet cone size $R$ in the C/A jet algorithm so that the Higgs tagging rates in each signal process were maximized compared to the SM background contributions. The cut flows to capture the kinematical features for the signal processes were applied thereafter. In particular, we optimize the $p_T$ cut to the tagged Higgs jets. Based on the analysis, the signal reaches for the $A\to hZ$ channel were obtained. The mass reach can be generally up to $\sim \mO(1)\,\TeV$ for the 2HDM-I with low-$t_\beta$ inputs at the HL LHC runs. The search mode is mostly interesting in the special $M_h = M_H =125\,\GeV$ degenerate scenario for the 2HDM-II case, both for the low-$t_\beta$ and large-$t_\beta$ regions. However, for the $M_h = 125\,\GeV$ scenario in the 2HDM-II, there exist stringent constraints on the alignment parameter $c_{\beta-\alpha}$ from the current global fit to the $125\,\GeV$ Higgs boson signal strengths. Therefore, this decay mode of $A\to hZ$ is highly suppressed in this case, unless the further results from the LHC measurements of the $125\,\GeV$ Higgs boson would modify the constraints significantly.

In more generic context with 2HDM setup as the low-energy description in the scalar sector, this decay mode of $A\to hZ$ exists. Studies to this decay mode for the CP-odd Higgs boson searches are of general interest in this sense for the future experiments. In particular, the analysis of the boosted Higgs jet from this channel can be similarly applied. As we have shown the sensitivity regions on the $(M_A\,,t_\beta)$ plane via this channel, the searches for the $A\to hZ$ mode can become complementary to the conventional search modes of $A\to \bar b b$ and $A\to \tau^+\tau^+$.

%###################################################################

\section*{Acknowledgments}

We would like to thank Yun Jiang for useful discussions. This work is partially supported by National Science Foundation of China (under grant No. 11275009, 11335007), the Tsinghua University Talent Fund (under grant No. 543481001), the Australian Research Council (under grant No. CE110001004). NC would like to thank the hospitality of the Center of High Energy Physics (CHEP) of Peking University when part of this work was prepared.


\begin{thebibliography}{999}

%\cite{Higgs:1964ia}
\bibitem{Higgs:1964ia} 
  P.~W.~Higgs,
  %``Broken symmetries, massless particles and gauge fields,''
  Phys.\ Lett.\  {\bf 12}, 132 (1964).
  %%CITATION = PHLTA,12,132;%%
  %2859 citations counted in INSPIRE as of 15 Oct 2014
  
  
  %\cite{Higgs:1964pj}
\bibitem{Higgs:1964pj} 
  P.~W.~Higgs,
  %``Broken Symmetries and the Masses of Gauge Bosons,''
  Phys.\ Rev.\ Lett.\  {\bf 13}, 508 (1964).
  %%CITATION = PRLTA,13,508;%%
  %2867 citations counted in INSPIRE as of 15 Oct 2014
  
  
  %\cite{Englert:1964et}
\bibitem{Englert:1964et} 
  F.~Englert and R.~Brout,
  %``Broken Symmetry and the Mass of Gauge Vector Mesons,''
  Phys.\ Rev.\ Lett.\  {\bf 13}, 321 (1964).
  %%CITATION = PRLTA,13,321;%%
  %2621 citations counted in INSPIRE as of 15 Oct 2014



%\cite{Peskin:2012we}
\bibitem{Peskin:2012we} 
  M.~E.~Peskin,
  %``Comparison of LHC and ILC Capabilities for Higgs Boson Coupling Measurements,''
  arXiv:1207.2516 [hep-ph].
  %%CITATION = ARXIV:1207.2516;%%
  %111 citations counted in INSPIRE as of 14 Oct 2014
  
  
  %\cite{Djouadi:2013yb}
\bibitem{Djouadi:2013yb} 
  A.~Djouadi, R.~M.~Godbole, B.~Mellado and K.~Mohan,
  %``Probing the spin-parity of the Higgs boson via jet kinematics in vector boson fusion,''
  Phys.\ Lett.\ B {\bf 723}, 307 (2013)
  [arXiv:1301.4965 [hep-ph]].
  %%CITATION = ARXIV:1301.4965;%%
  %42 citations counted in INSPIRE as of 14 Oct 2014


%\cite{Curtin:2013fra}
\bibitem{Curtin:2013fra} 
  D.~Curtin, R.~Essig, S.~Gori, P.~Jaiswal, A.~Katz, T.~Liu, Z.~Liu and D.~McKeen {\it et al.},
  %``Exotic Decays of the 125 GeV Higgs Boson,''
  arXiv:1312.4992 [hep-ph].
  %%CITATION = ARXIV:1312.4992;%%
  %22 citations counted in INSPIRE as of 14 Oct 2014


%\cite{Dimopoulos:1981zb}
\bibitem{Dimopoulos:1981zb} 
  S.~Dimopoulos and H.~Georgi,
  %``Softly Broken Supersymmetry and SU(5),''
  Nucl.\ Phys.\ B {\bf 193}, 150 (1981).
  %%CITATION = NUPHA,B193,150;%%
  %2089 citations counted in INSPIRE as of 14 Oct 2014
  
  
  %\cite{Chacko:2005vw}
\bibitem{Chacko:2005vw} 
  Z.~Chacko, Y.~Nomura, M.~Papucci and G.~Perez,
  %``Natural little hierarchy from a partially goldstone twin Higgs,''
  JHEP {\bf 0601}, 126 (2006)
  [hep-ph/0510273].
  %%CITATION = HEP-PH/0510273;%%
  %63 citations counted in INSPIRE as of 14 Oct 2014
  
  
  %\cite{Mrazek:2011iu}
\bibitem{Mrazek:2011iu} 
  J.~Mrazek, A.~Pomarol, R.~Rattazzi, M.~Redi, J.~Serra and A.~Wulzer,
  %``The Other Natural Two Higgs Doublet Model,''
  Nucl.\ Phys.\ B {\bf 853}, 1 (2011)
  [arXiv:1105.5403 [hep-ph]].
  %%CITATION = ARXIV:1105.5403;%%
  %64 citations counted in INSPIRE as of 14 Oct 2014
  
  %\cite{Branco:2011iw}
\bibitem{Branco:2011iw} 
  G.~C.~Branco, P.~M.~Ferreira, L.~Lavoura, M.~N.~Rebelo, M.~Sher and J.~P.~Silva,
  %``Theory and phenomenology of two-Higgs-doublet models,''
  Phys.\ Rept.\  {\bf 516}, 1 (2012)
  [arXiv:1106.0034 [hep-ph]].
  %%CITATION = ARXIV:1106.0034;%%
  %153 citations counted in INSPIRE as of 19 May 2013
  
  
  %\cite{Craig:2012vn}
\bibitem{Craig:2012vn} 
  N.~Craig and S.~Thomas,
  %``Exclusive Signals of an Extended Higgs Sector,''
  JHEP {\bf 1211}, 083 (2012)
  [arXiv:1207.4835 [hep-ph]].
  %%CITATION = ARXIV:1207.4835;%%
  %50 citations counted in INSPIRE as of 11 Oct 2014
  
  
  %\cite{Craig:2012pu}
\bibitem{Craig:2012pu} 
  N.~Craig, J.~A.~Evans, R.~Gray, C.~Kilic, M.~Park, S.~Somalwar and S.~Thomas,
  %``Multi-Lepton Signals of Multiple Higgs Bosons,''
  JHEP {\bf 1302}, 033 (2013)
  [arXiv:1210.0559 [hep-ph]].
  %%CITATION = ARXIV:1210.0559;%%
  %21 citations counted in INSPIRE as of 11 Oct 2014
  
      %\cite{Coleppa:2013dya}
\bibitem{Coleppa:2013dya} 
  B.~Coleppa, F.~Kling and S.~Su,
  %``Constraining Type II 2HDM in Light of LHC Higgs Searches,''
  arXiv:1305.0002 [hep-ph].
  %%CITATION = ARXIV:1305.0002;%%
  %6 citations counted in INSPIRE as of 23 May 2013
  
  
  %\cite{Craig:2013hca}
\bibitem{Craig:2013hca} 
  N.~Craig, J.~Galloway and S.~Thomas,
  %``Searching for Signs of the Second Higgs Doublet,''
  arXiv:1305.2424 [hep-ph].
  %%CITATION = ARXIV:1305.2424;%%
  %65 citations counted in INSPIRE as of 08 Oct 2014
  

  

  %\cite{Coleppa:2013xfa}
\bibitem{Coleppa:2013xfa} 
  B.~Coleppa, F.~Kling and S.~Su,
  %``Exotic Higgs Decay via AZ/HZ Channel: a Snowmass Whitepaper,''
  arXiv:1308.6201 [hep-ph].
  %%CITATION = ARXIV:1308.6201;%%
  %1 citations counted in INSPIRE as of 22 Sep 2013


  %\cite{Carena:2013ooa}
\bibitem{Carena:2013ooa} 
  M.~Carena, I.~Low, N.~R.~Shah and C.~E.~M.~Wagner,
  %``Impersonating the Standard Model Higgs Boson: Alignment without Decoupling,''
  arXiv:1310.2248 [hep-ph].
  %%CITATION = ARXIV:1310.2248;%%
  
 
  %\cite{Chen:2013emb}
\bibitem{Chen:2013emb} 
  N.~Chen, C.~Du, Y.~Fang and L.~C.~Lü,
  %``LHC Searches for The Heavy Higgs Boson via Two B Jets plus Diphoton,''
  Phys.\ Rev.\ D {\bf 89}, 115006 (2014)
  [arXiv:1312.7212 [hep-ph]].
  %%CITATION = ARXIV:1312.7212;%%
  %8 citations counted in INSPIRE as of 11 Oct 2014
  
  
  
  %\cite{Baglio:2014nea}
\bibitem{Baglio:2014nea} 
  J.~Baglio, O.~Eberhardt, U.~Nierste and M.~Wiebusch,
  %``Benchmarks for Higgs Pair Production and Heavy Higgs Searches in the Two-Higgs-Doublet Model of Type II,''
  Phys.\ Rev.\ D {\bf 90}, 015008 (2014)
  [arXiv:1403.1264 [hep-ph]].
  %%CITATION = ARXIV:1403.1264;%%
  %20 citations counted in INSPIRE as of 14 Oct 2014
  
  
  %\cite{Coleppa:2014hxa}
\bibitem{Coleppa:2014hxa} 
  B.~Coleppa, F.~Kling and S.~Su,
  %``Exotic Decays Of A Heavy Neutral Higgs Through HZ/AZ Channel,''
  JHEP {\bf 1409}, 161 (2014)
  [arXiv:1404.1922 [hep-ph]].
  %%CITATION = ARXIV:1404.1922;%%
  %7 citations counted in INSPIRE as of 14 Oct 2014
  
  
  %\cite{Dumont:2014wha}
\bibitem{Dumont:2014wha} 
  B.~Dumont, J.~F.~Gunion, Y.~Jiang and S.~Kraml,
  %``Constraints on and future prospects for Two-Higgs-Doublet Models in light of the LHC Higgs signal,''
  Phys.\ Rev.\ D {\bf 90}, 035021 (2014)
  [arXiv:1405.3584 [hep-ph]].
  %%CITATION = ARXIV:1405.3584;%%
  %18 citations counted in INSPIRE as of 14 Oct 2014
  
  
  %\cite{Dorsch:2014qja}
\bibitem{Dorsch:2014qja} 
  G.~C.~Dorsch, S.~Huber, K.~Mimasu and J.~M.~No,
  %``Echoes of the Electroweak Phase Transition: Discovering a second Higgs doublet through $A_0 \rightarrow H_0 Z$,''
  arXiv:1405.5537 [hep-ph].
  %%CITATION = ARXIV:1405.5537;%%
  %1 citations counted in INSPIRE as of 14 Oct 2014
  
  
  
  %\cite{Hespel:2014sla}
\bibitem{Hespel:2014sla} 
  B.~Hespel, D.~Lopez-Val and E.~Vryonidou,
  %``Higgs pair production via gluon fusion in the Two-Higgs-Doublet Model,''
  JHEP {\bf 1409}, 124 (2014)
  [arXiv:1407.0281 [hep-ph]].
  %%CITATION = ARXIV:1407.0281;%%
  %4 citations counted in INSPIRE as of 14 Oct 2014
  
  
  %\cite{Barger:2014qva}
\bibitem{Barger:2014qva} 
  V.~Barger, L.~L.~Everett, C.~B.~Jackson, A.~D.~Peterson and G.~Shaughnessy,
  %``Measuring the 2HDM Scalar Potential at LHC14,''
  arXiv:1408.2525 [hep-ph].
  %%CITATION = ARXIV:1408.2525;%%
  %2 citations counted in INSPIRE as of 14 Oct 2014
  
  
  
  
  %\cite{Fontes:2014xva}
\bibitem{Fontes:2014xva} 
  D.~Fontes, J.~C.~Romão and J.~P.~Silva,
  %``$h \rightarrow Z \gamma$ in the complex two Higgs doublet model,''
  arXiv:1408.2534 [hep-ph].
  %%CITATION = ARXIV:1408.2534;%%
  %1 citations counted in INSPIRE as of 14 Oct 2014
  
  
  
  %\cite{Coleppa:2014cca}
\bibitem{Coleppa:2014cca} 
  B.~Coleppa, F.~Kling and S.~Su,
  %``Charged Higgs Search via $AW^\pm/HW^\pm$ Channel,''
  arXiv:1408.4119 [hep-ph].
  %%CITATION = ARXIV:1408.4119;%%
  
  
    
    
    %\cite{Grzadkowski:2014ada}
\bibitem{Grzadkowski:2014ada} 
  B.~Grzadkowski, O.~M.~Ogreid and P.~Osland,
  %``Measuring CP violation in Two-Higgs-Doublet models in light of the LHC Higgs data,''
  arXiv:1409.7265 [hep-ph].
  %%CITATION = ARXIV:1409.7265;%%

    
    
  
  %\cite{Gunion:2012he}
\bibitem{Gunion:2012he} 
  J.~F.~Gunion, Y.~Jiang and S.~Kraml,
  %``Diagnosing Degenerate Higgs Bosons at 125 GeV,''
  Phys.\ Rev.\ Lett.\  {\bf 110}, 051801 (2013)
  [arXiv:1208.1817 [hep-ph]].
  %%CITATION = ARXIV:1208.1817;%%
  %39 citations counted in INSPIRE as of 11 Oct 2014
  
  %\cite{Aaltonen:2012zh}
\bibitem{Aaltonen:2012zh} 
  T.~Aaltonen {\it et al.}  [CDF and D0 Collaborations],
  %``Search for Neutral Higgs Bosons in Events with Multiple Bottom Quarks at the Tevatron,''
  Phys.\ Rev.\ D {\bf 86}, 091101 (2012)
  [arXiv:1207.2757 [hep-ex]].
  %%CITATION = ARXIV:1207.2757;%%
  %14 citations counted in INSPIRE as of 12 Oct 2014
  
  
  %\cite{Chatrchyan:2013qga}
\bibitem{Chatrchyan:2013qga} 
  S.~Chatrchyan {\it et al.}  [CMS Collaboration],
  %``Search for a Higgs boson decaying into a b-quark pair and produced in association with b quarks in proton-proton collisions at 7 TeV,''
  Phys.\ Lett.\ B {\bf 722}, 207 (2013)
  [arXiv:1302.2892 [hep-ex]].
  %%CITATION = ARXIV:1302.2892;%%
  %30 citations counted in INSPIRE as of 12 Oct 2014
  
  %\cite{Schael:2006cr}
\bibitem{Schael:2006cr} 
  S.~Schael {\it et al.}  [ALEPH and DELPHI and L3 and OPAL and LEP Working Group for Higgs Boson Searches Collaborations],
  %``Search for neutral MSSM Higgs bosons at LEP,''
  Eur.\ Phys.\ J.\ C {\bf 47}, 547 (2006)
  [hep-ex/0602042].
  %%CITATION = HEP-EX/0602042;%%
  %674 citations counted in INSPIRE as of 12 Oct 2014
  
  
  
  %\cite{Abazov:2008hu}
\bibitem{Abazov:2008hu} 
  V.~M.~Abazov {\it et al.}  [D0 Collaboration],
  %``Search for Higgs bosons decaying to $\tau$ pairs in $p \bar{p}$ collisions with the D0 detector,''
  Phys.\ Rev.\ Lett.\  {\bf 101}, 071804 (2008)
  [arXiv:0805.2491 [hep-ex]].
  %%CITATION = ARXIV:0805.2491;%%
  %60 citations counted in INSPIRE as of 12 Oct 2014
  
  %\cite{Aaltonen:2009vf}
\bibitem{Aaltonen:2009vf} 
  T.~Aaltonen {\it et al.}  [CDF Collaboration],
  %``Search for Higgs bosons predicted in two-Higgs-doublet models via decays to tau lepton pairs in 1.96-TeV p anti-p collisions,''
  Phys.\ Rev.\ Lett.\  {\bf 103}, 201801 (2009)
  [arXiv:0906.1014 [hep-ex]].
  %%CITATION = ARXIV:0906.1014;%%
  %51 citations counted in INSPIRE as of 12 Oct 2014
  
  
  %\cite{Abazov:2011up}
\bibitem{Abazov:2011up} 
  V.~M.~Abazov {\it et al.}  [D0 Collaboration],
  %``Search for Higgs bosons of the minimal supersymmetric standard model in $p\bar{p}$ collisions at $\sqrt(s)=1.96$ TeV,''
  Phys.\ Lett.\ B {\bf 710}, 569 (2012)
  [arXiv:1112.5431 [hep-ex]].
  %%CITATION = ARXIV:1112.5431;%%
  %13 citations counted in INSPIRE as of 12 Oct 2014
  
  %\cite{Chatrchyan:2012vp}
\bibitem{Chatrchyan:2012vp} 
  S.~Chatrchyan {\it et al.}  [CMS Collaboration],
  %``Search for neutral Higgs bosons decaying to $\tau$ pairs in $pp$ collisions at $\sqrt{s}=7$ TeV,''
  Phys.\ Lett.\ B {\bf 713}, 68 (2012)
  [arXiv:1202.4083 [hep-ex]].
  %%CITATION = ARXIV:1202.4083;%%
  %180 citations counted in INSPIRE as of 12 Oct 2014
  
  
  %\cite{Aad:2012cfr}
\bibitem{Aad:2012cfr} 
  G.~Aad {\it et al.}  [ATLAS Collaboration],
  %``Search for the neutral Higgs bosons of the Minimal Supersymmetric Standard Model in $pp$ collisions at $\sqrt{s}=7$ TeV with the ATLAS detector,''
  JHEP {\bf 1302}, 095 (2013)
  [arXiv:1211.6956 [hep-ex]].
  %%CITATION = ARXIV:1211.6956;%%
  %58 citations counted in INSPIRE as of 12 Oct 2014

  
 
  %\cite{Aad:2014vgg}
\bibitem{Aad:2014vgg} 
  G.~Aad {\it et al.}  [ ATLAS Collaboration],
  %``Search for neutral Higgs bosons of the minimal supersymmetric standard model in pp collisions at $\sqrt{s}$ = 8 TeV with the ATLAS detector,''
  arXiv:1409.6064 [hep-ex].
  %%CITATION = ARXIV:1409.6064;%%
  %1 citations counted in INSPIRE as of 11 Oct 2014
  
  
  
  %\cite{CMS:2013eua}
\bibitem{CMS:2013eua} 
  CMS Collaboration [CMS Collaboration],
  %``Search for extended Higgs sectors in the H to hh and A to Zh channels in sqrt(s) = 8 TeV pp collisions with multileptons and photons final states,''
  CMS-PAS-HIG-13-025.
  %%CITATION = CMS-PAS-HIG-13-025;%%
  %21 citations counted in INSPIRE as of 20 Oct 2014
  
  
 %\cite{Butterworth:2008iy}
\bibitem{Butterworth:2008iy} 
  J.~M.~Butterworth, A.~R.~Davison, M.~Rubin and G.~P.~Salam,
  %``Jet substructure as a new Higgs search channel at the LHC,''
  Phys.\ Rev.\ Lett.\  {\bf 100}, 242001 (2008)
  [arXiv:0802.2470 [hep-ph]].
  %%CITATION = ARXIV:0802.2470;%%
  %442 citations counted in INSPIRE as of 12 Oct 2014
  
  
  
  %\cite{Butterworth:2008tr}
\bibitem{Butterworth:2008tr} 
  J.~M.~Butterworth, A.~R.~Davison, M.~Rubin and G.~P.~Salam,
  %``Jet substructure as a new Higgs search channel at the LHC,''
  arXiv:0810.0409 [hep-ph].
  %%CITATION = ARXIV:0810.0409;%%
  %13 citations counted in INSPIRE as of 15 Oct 2014
  
  %\cite{Carena:2010ev}
\bibitem{Carena:2010ev} 
  M.~Carena, P.~Draper, S.~Heinemeyer, T.~Liu, C.~E.~M.~Wagner and G.~Weiglein,
  %``Probing the Higgs Sector of High-Scale SUSY-Breaking Models at the Tevatron,''
  Phys.\ Rev.\ D {\bf 83}, 055007 (2011)
  [arXiv:1011.5304 [hep-ph]].
  %%CITATION = ARXIV:1011.5304;%%
  %4 citations counted in INSPIRE as of 15 Oct 2014
  
  
    %\cite{Yang:2011jk}
\bibitem{Yang:2011jk} 
  S.~Yang and Q.~S.~Yan,
  %``Searching for Heavy Charged Higgs Boson with Jet Substructure at the LHC,''
  JHEP {\bf 1202}, 074 (2012)
  [arXiv:1111.4530 [hep-ph]].
  %%CITATION = ARXIV:1111.4530;%%
  %17 citations counted in INSPIRE as of 08 Oct 2014
  
  
  %\cite{Kang:2013rj}
\bibitem{Kang:2013rj} 
  Z.~Kang, J.~Li, T.~Li, D.~Liu and J.~Shu,
  %``Probing the CP-even Higgs sector via $H_3$ → $H_2H_1$ in the natural next-to-minimal supersymmetric standard model,''
  Phys.\ Rev.\ D {\bf 88}, no. 1, 015006 (2013)
  [arXiv:1301.0453 [hep-ph]].
  %%CITATION = ARXIV:1301.0453;%%
  %22 citations counted in INSPIRE as of 08 Oct 2014

  
  %\cite{Plehn:2009rk}
\bibitem{Plehn:2009rk} 
  T.~Plehn, G.~P.~Salam and M.~Spannowsky,
  %``Fat Jets for a Light Higgs,''
  Phys.\ Rev.\ Lett.\  {\bf 104}, 111801 (2010)
  [arXiv:0910.5472 [hep-ph]].
  %%CITATION = ARXIV:0910.5472;%%
  %186 citations counted in INSPIRE as of 15 Oct 2014
  
  
  
  %\cite{Godbole:2013saa}
\bibitem{Godbole:2013saa} 
  R.~Godbole, D.~J.~Miller, K.~Mohan and C.~D.~White,
  %``Boosting Higgs CP properties via $VH$ Production at the Large Hadron Collider,''
  Phys.\ Lett.\ B {\bf 730}, 275 (2014)
  [arXiv:1306.2573 [hep-ph]].
  %%CITATION = ARXIV:1306.2573;%%
  %13 citations counted in INSPIRE as of 22 Oct 2014
  
%\cite{Godbole:2014cfa}
\bibitem{Godbole:2014cfa} 
  R.~M.~Godbole, D.~J.~Miller, K.~A.~Mohan and C.~D.~White,
  %``Jet substructure and probes of CP violation in Vh production,''
  arXiv:1409.5449 [hep-ph].
  %%CITATION = ARXIV:1409.5449;%%
  
  
  %\cite{Dokshitzer:1997in}
\bibitem{Dokshitzer:1997in} 
  Y.~L.~Dokshitzer, G.~D.~Leder, S.~Moretti and B.~R.~Webber,
  %``Better jet clustering algorithms,''
  JHEP {\bf 9708}, 001 (1997)
  [hep-ph/9707323].
  %%CITATION = HEP-PH/9707323;%%
  %461 citations counted in INSPIRE as of 13 Oct 2014
  
  
  %\cite{Wobisch:1998wt}
\bibitem{Wobisch:1998wt} 
  M.~Wobisch and T.~Wengler,
  %``Hadronization corrections to jet cross-sections in deep inelastic scattering,''
  In *Hamburg 1998/1999, Monte Carlo generators for HERA physics* 270-279
  [hep-ph/9907280].
  %%CITATION = HEP-PH/9907280;%%
  %247 citations counted in INSPIRE as of 13 Oct 2014
  
  
  
   %\cite{Barger:2013ofa}
\bibitem{Barger:2013ofa} 
  V.~Barger, L.~L.~Everett, H.~E.~Logan and G.~Shaughnessy,
  %``Scrutinizing the 125 GeV Higgs boson in two Higgs doublet models at the LHC, ILC, and Muon Collider,''
  Phys.\ Rev.\ D {\bf 88}, no. 11, 115003 (2013)
  [arXiv:1308.0052 [hep-ph]].
  %%CITATION = ARXIV:1308.0052;%%
  %31 citations counted in INSPIRE as of 08 Oct 2014
  
  
  %\cite{Djouadi:2005gi}
\bibitem{Djouadi:2005gi} 
  A.~Djouadi,
  %``The Anatomy of electro-weak symmetry breaking. I: The Higgs boson in the standard model,''
  Phys.\ Rept.\  {\bf 457}, 1 (2008)
  [hep-ph/0503172].
  %%CITATION = HEP-PH/0503172;%%
  
  
  
  %\cite{Djouadi:2005gj}
\bibitem{Djouadi:2005gj} 
  A.~Djouadi,
  %``The Anatomy of electro-weak symmetry breaking. II. The Higgs bosons in the minimal supersymmetric model,''
  Phys.\ Rept.\  {\bf 459}, 1 (2008)
  [hep-ph/0503173].
  %%CITATION = HEP-PH/0503173;%%
  

  
  
  %\cite{Harlander:2012pb}
\bibitem{Harlander:2012pb} 
  R.~V.~Harlander, S.~Liebler and H.~Mantler,
  %``SusHi: A program for the calculation of Higgs production in gluon fusion and bottom-quark annihilation in the Standard Model and the MSSM,''
  Comput.\ Phys.\ Commun.\  {\bf 184}, 1605 (2013)
  [arXiv:1212.3249 [hep-ph]].
  %%CITATION = ARXIV:1212.3249;%%
  %38 citations counted in INSPIRE as of 08 Oct 2014

   %\cite{Eriksson:2009ws}
\bibitem{Eriksson:2009ws} 
  D.~Eriksson, J.~Rathsman and O.~Stal,
  %``2HDMC: Two-Higgs-Doublet Model Calculator Physics and Manual,''
  Comput.\ Phys.\ Commun.\  {\bf 181}, 189 (2010)
  [arXiv:0902.0851 [hep-ph]].
  %%CITATION = ARXIV:0902.0851;%%
  %33 citations counted in INSPIRE as of 19 Jun 2013

 %\cite{Ahrens:2011px}
\bibitem{Ahrens:2011px} 
  V.~Ahrens, A.~Ferroglia, M.~Neubert, B.~D.~Pecjak and L.~L.~Yang,
  %``Precision predictions for the t+t(bar) production cross section at hadron colliders,''
  Phys.\ Lett.\ B {\bf 703}, 135 (2011)
  [arXiv:1105.5824 [hep-ph]].
  %%CITATION = ARXIV:1105.5824;%%
  %83 citations counted in INSPIRE as of 27 Oct 2014 
  
  %\cite{Cordero:2009kv}
\bibitem{Cordero:2009kv} 
  F.~Febres Cordero, L.~Reina and D.~Wackeroth,
  %``W- and Z-boson production with a massive bottom-quark pair at the Large Hadron Collider,''
  Phys.\ Rev.\ D {\bf 80}, 034015 (2009)
  [arXiv:0906.1923 [hep-ph]].
  %%CITATION = ARXIV:0906.1923;%%
  %30 citations counted in INSPIRE as of 27 Oct 2014
  
  
  %\cite{Dittmaier:2011ti}
\bibitem{Dittmaier:2011ti} 
  S.~Dittmaier {\it et al.}  [LHC Higgs Cross Section Working Group Collaboration],
  %``Handbook of LHC Higgs Cross Sections: 1. Inclusive Observables,''
  arXiv:1101.0593 [hep-ph].
  %%CITATION = ARXIV:1101.0593;%%
  %786 citations counted in INSPIRE as of 27 Oct 2014
  
  %\cite{Campbell:2011bn}
\bibitem{Campbell:2011bn} 
  J.~M.~Campbell, R.~K.~Ellis and C.~Williams,
  %``Vector boson pair production at the LHC,''
  JHEP {\bf 1107}, 018 (2011)
  [arXiv:1105.0020 [hep-ph]].
  %%CITATION = ARXIV:1105.0020;%%
  %390 citations counted in INSPIRE as of 27 Oct 2014
  
  
        %\cite{ATLAS:2013079}
\bibitem{ATLAS:2013079} 
  [ATLAS Collaboration],
  ATLAS-CONF-2013-079.
  %%CITATION = ATLAS-CONF-2013-079;%%  
  
  
   %\cite{ATLAS:2012aoa}
\bibitem{ATLAS:2012aoa} 
  [ATLAS Collaboration],
  %``Measuring the b-tag efficiency in a top-pair sample with 4.7 fb^-1 of data from the ATLAS detector,''
  ATLAS-CONF-2012-097.
  %%CITATION = ATLAS-CONF-2012-097;%%
  %12 citations counted in INSPIRE as of 05 Jun 2013

  
    %\cite{ATLAS:2012lma}
\bibitem{ATLAS:2012lma} 
  [ATLAS Collaboration],
  %``Measurement of the Mistag Rate with 5 fb$^{−1}$ of Data Collected by the ATLAS Detector,''
  ATLAS-CONF-2012-040.
  %%CITATION = ATLAS-CONF-2012-040;%%
  %53 citations counted in INSPIRE as of 05 Jun 2013

  

  %\cite{Christensen:2008py}
\bibitem{Christensen:2008py} 
  N.~D.~Christensen and C.~Duhr,
  %``FeynRules - Feynman rules made easy,''
  Comput.\ Phys.\ Commun.\  {\bf 180}, 1614 (2009)
  [arXiv:0806.4194 [hep-ph]].
  %%CITATION = ARXIV:0806.4194;%%
  %202 citations counted in INSPIRE as of 06 Jun 2013
  
  

  
  
  %\cite{Alwall:2014hca}
\bibitem{Alwall:2014hca} 
  J.~Alwall, R.~Frederix, S.~Frixione, V.~Hirschi, F.~Maltoni, O.~Mattelaer, H.-S.~Shao and T.~Stelzer {\it et al.},
  %``The automated computation of tree-level and next-to-leading order differential cross sections, and their matching to parton shower simulations,''
  JHEP {\bf 1407}, 079 (2014)
  [arXiv:1405.0301 [hep-ph]].
  %%CITATION = ARXIV:1405.0301;%%
  %104 citations counted in INSPIRE as of 27 Oct 2014
  
  
  
  
  %\cite{Sjostrand:2006za}
\bibitem{Sjostrand:2006za} 
  T.~Sjostrand, S.~Mrenna and P.~Z.~Skands,
  %``PYTHIA 6.4 Physics and Manual,''
  JHEP {\bf 0605}, 026 (2006)
  [hep-ph/0603175].
  %%CITATION = HEP-PH/0603175;%%
  %4289 citations counted in INSPIRE as of 12 Aug 2013
  

    %\cite{deFavereau:2013fsa}
\bibitem{deFavereau:2013fsa} 
  J.~de Favereau, C.~Delaere, P.~Demin, A.~Giammanco, V.~Lemaître, A.~Mertens and M.~Selvaggi,
  %``DELPHES 3, A modular framework for fast simulation of a generic collider experiment,''
  arXiv:1307.6346 [hep-ex].
  %%CITATION = ARXIV:1307.6346;%%
  %6 citations counted in INSPIRE as of 12 Aug 2013
  
  
  
  %\cite{Cacciari:2011ma}
\bibitem{Cacciari:2011ma} 
  M.~Cacciari, G.~P.~Salam and G.~Soyez,
  %``FastJet User Manual,''
  Eur.\ Phys.\ J.\ C {\bf 72}, 1896 (2012)
  [arXiv:1111.6097 [hep-ph]].
  %%CITATION = ARXIV:1111.6097;%%
  %642 citations counted in INSPIRE as of 13 Oct 2014
  
  
  
  
  
   
 
\end{thebibliography}
\end{document}